\documentclass[twocolumn,showpacs,preprintnumbers,amsmath,amssymb]{revtex4}

\usepackage{epsf}
\usepackage{rotating}

\usepackage{graphicx,epsfig}
\usepackage{dcolumn}
\usepackage{bm}
\usepackage{color}

\def\cN{{\cal N}}

\def\cH{{\cal H}}
\def\cM{{\cal M}}
\def\cS{{\cal S}}
\def\cQ{{\cal Q}}

\def\flow{{I}}
\newcommand{\req}[1]{Eq.~(\ref{#1})}
\newcommand{\avg}[1]{\langle #1\rangle}

\newcommand{\fig}[1]{Fig.~\ref{#1}}

\DeclareMathOperator*{\argmin}{{\rm argmin}}	

\newcommand{\cut}[1]{{}}
\newcommand{\etal}[1]{\emph{~et al.}}

\newcommand{\hilm}{h_{i\to l}^{(\sigma)}}
\newcommand{\gilm}{g_{i\to l}^{(\sigma)}}
\newcommand{\vhil}{\vec{h}_{i\to l}}
\newcommand{\vgil}{\vec{g}_{i\to l}}
\newcommand{\vhji}{\vec{h}_{j\to i}}
\newcommand{\vgji}{\vec{g}_{j\to i}}
\newcommand{\hil}{h_{i\to l}}
\newcommand{\gil}{g_{i\to l}}
\newcommand{\hji}{h_{j\to i}}
\newcommand{\gji}{g_{j\to i}}
\newcommand{\hjinl}{\hat{h}_{i\backslash l}}
\newcommand{\gjinl}{\hat{g}_{i\backslash l}}
\newcommand{\Isjinl}{\hat{I}^*_{i\backslash l}}
\newcommand{\Isil}{\flow^*_{il}}
\newcommand{\Isji}{\flow^*_{ji}}
\newcommand{\fa}{f_{\rm a}}
\newcommand{\fcon}{f_{\rm con}}
\newcommand{\Non}{N_{\rm a}}

\newcommand{\fai}{f_{\rm a-i}}
\newcommand{\faa}{f_{\rm a-a}}
\newcommand{\fii}{f_{\rm i-i}}
\newcommand{\fan}{f_{\rm AN}}
\newcommand{\fon}{f_{\rm ON}}

\begin{document}

\preprint{}

\title[Title]
{Coverage versus Supply Cost in Facility Location: Physics of Frustrated Spin Systems
}
\author{Chi Ho Yeung$^{1,2,3}$, K. Y. Michael Wong$^2$ and Bo Li$^2$}
\affiliation{$^1$The Nonlinearity and Complexity Research Group, Aston University, Birmingham B4 7ET, United Kingdom
\\
$^2$Department of Physics, The Hong Kong University of Science and Technology, Hong Kong
\\
$^3$Department of Science and Environmental Studies, The Hong Kong Institute of Education, Tai Po, Hong Kong
}

\date{\today}

\begin{abstract}
A comprehensive coverage is crucial for communication, supply and transportation networks, yet it is limited by the  requirement of extensive infrastructure and heavy energy consumption. Here we draw an analogy between spins in antiferromagnet and outlets in supply networks, and apply techniques from the studies of disordered systems to elucidate the effects of balancing the coverage and supply costs on the network behavior. A readily applicable, coverage optimization algorithm is derived. Simulation results show that magnetized and antiferromagnetic domains emerge and coexist to balance the need for coverage and energy saving. The scaling of parameters with system size agrees with the continuum approximation in two dimensions and the tree approximation in random graphs. Due to frustration caused by the competition between coverage and supply cost, a transition between easy and hard computation regimes is observed. We further suggest a local expansion approach to greatly simplify the message updates which shed light on simplifications in other problems.
\end{abstract}

\pacs{89.75.Hc, 02.50.-r, 05.20.-y, 89.20.-a}

\maketitle


The effectiveness of communication and supply networks relies crucially on the extent of network coverage. For instance, networks of fire sensors, surveillance video cameras, local weather monitors require a comprehensive coverage of a geographical region to guard safety and security or collect extensive information~\cite{rangwala06}. A comprehensive coverage is essential in transportation~\cite{megiddo81} and water distribution networks~\cite{vasan10}. Likewise, the locations of supermarket branches, teller machines or restaurant outlets determine market share, and hence business success~\cite{ghosh83, berman90}. Governments set up public facilities such as schools and clinics to cover their administrative regions so as to uphold the well-being of the community~\cite{revelle77}. 

On the other hand, it is costly to extend and maintain coverage. Sensor networks usually consume a lot of transmission power for broad area coverage, which is a problem for maintaining the life span of the sensor networks given the limited battery size of the sensors~\cite{alkaraki04, frey09}. The logistics supply for networks of facilities requires abundant resources as well as extensive infrastructure. In military history, overstretching the supply lines increased the vulnerability of military operations, as commonly believed in the Russian campaigns of Napoelon and Hitler~\cite{seward88}. These create practical concerns on coverage expansion, and lead to a trade-off between supply costs and service coverage.

To balance the need for coverage and the cost of maintaining supplies, efforts have been devoted in various disciplines. For instance, deployment of sensors in sensor networks has been intensely studied by engineers~\cite{clouquer02, zou03}; retail strategies and optimal branch allocation are studied in operations research~\cite{ghosh83, berman90}; airline destinations are frequently revised by airline management. However, many of the above problems remain heuristically tackled, which result in suboptimal solutions. In addition, centralized techniques such as linear programming are usually employed~\cite{plastria01}. These work for problems with a global optimizer, but not for distributive applications such as the distribution of P2P file storage on the Internet ~\cite{wong08}.

This competition shares similarities with frustrated spin systems in statistical physics~\cite{vannimenus77}. To see the correspondence, we note that local variables in a spin model are represented by ``spins" with binary  states (but can be generalized to multiple  states) and correspond to whether a location is selected for facility location or not. The propensity to increase coverage can be modeled by antiferromagnetic interactions.  At the same time, we introduce  a transportation cost from a warehouse to each selected node. This will result in a regional bias of selected nodes in the neighborhood of the warehouse, in competition with antiferromagnetism. Our formulation is also relevant to sensor networks where distributed sensors communicate with a base station. In the context of spin models, the transportation cost creates a long-range correlation between spins, a new feature absent in conventional antiferromagnet models. 

In this paper, we elucidate the different patterns resultant from the interplay between these factors. We will find the coexistence of magnetized, and antiferromagnetic or glassy domains. This multiplicity of possible states is a consequence of frustration, and leads to the appearance of many suboptimal solutions, and interferes severely with the search for the optimal solution. By applying techniques in the studies of disordered systems~\cite{mezard87, nishimori01}, we derive a readily applicable distributed algorithm capable of optimizing real instances.  In addition, the comparison between square lattices and random regular graphs lead to important implications on their coverage efficiency and cost scalability.

Other than the above implications, we have extended the local expansion approach in~\cite{wong06} to intensely simplify the suggested message passing procedures.   This algorithm sheds light on similar simplifications in other physical and optimization problems. A simple example of convergence to metastable states also reveals the limitation of fundamental message passing algorithms.

\section{The Model}

Specifically, we consider a network of $N$ nodes; each node $i=1,\cdots,N$ is connected to a set of $k_i$ neighbors denoted by $\cN_i$. We denote $s_i = 1, -1$ when node $i$ is \emph{active} or \emph{idle} respectively. An active node can be regarded as an outlet, a service station or a sensor which serves or monitors the surrounding area. In addition, each of the active nodes establishes an individual communication path to a \emph{central termina}l, labeled $T$, where commodities are stored or information is processed. We denote the total flow on the edge from $i$ to $j$ as $I_{ij} (= -I_{ji})$, which is the sum of communication paths passing the edge $(ij)$. We then minimize the Hamiltonian $H$ in the space of $s_i$ and $I_{ij}$, given by
\begin{align}
\label{eq_H}
\cH = J\sum_{(ij)}s_is_j+U\sum_{i=1}^N \frac{1-s_i}{2}+\sum_{(ij)}\phi(I_{ij}).
\end{align}
In the first term, a positive $J$ encourages neighbors to be in opposite states. In  supply networks, $J$ may correspond to the \emph{outlet redundancy}, namely, the redundant operational cost due to the duplication of service in neighboring nodes. It tends to spread the outlets and hence increase the coverage. The second term represents the loss in revenue due to nodes being uncovered. The \emph{idle penalty} $U$  may correspond to unsatisfied demand, such that a large $U$ encourages outlet installation. The third term is the \emph{supply cost}. Although our analytic and algorithmic results are generic to the form of $\phi(x)$, we assume $\phi(x) =x^2$ in the present study. Considering situations that a single path is established between individual active nodes and the terminal, we restrict the current $I_{ij}$ on all edges $(ij)$ to be integers ~\cite{yeung12}, i.e.
\begin{align}
\label{eq_integer}
I_{ij}\in \mathbb{Z},\qquad\forall (ij).
\end{align}
The variables are subject to the following constraint to conserve the flow of resources,
\begin{align}
\label{eq_conserve}
\sum_{j\in \cN_i}I_{ji} = \frac{1+s_i}{2}.
\end{align}
In this paper, we will mainly study the interplay between parameter $J$ and $U$ and the optimal coverage state. We remark that it is straightforward to generalize the model to accommodate heterogeneous on-site demand through multiplying the right hand side of \req{eq_conserve} by the demand of node $i$.

The above set-up is analogous to spin models. The first and second terms of $H$ correspond to an anti-ferromagnetic interaction and an external magnetic field respectively. The last term is ubiquitous to communication networks~\cite{wong06}. If one fixes the active or inactive states of the nodes and optimize the supply energy in the space of $I_{ij}$, one would obtain an effective long-range antiferromagnetic interaction between the spins, since activating two nodes which share part of their paths to the terminal will increase the nonlinear total supply cost. The interplay of these three interactions would lead to new physical phenomena.

\begin{figure}
\leftline{\hspace{1.2cm}(a)}
\vspace{-0.3cm}
\centerline{\epsfig{figure=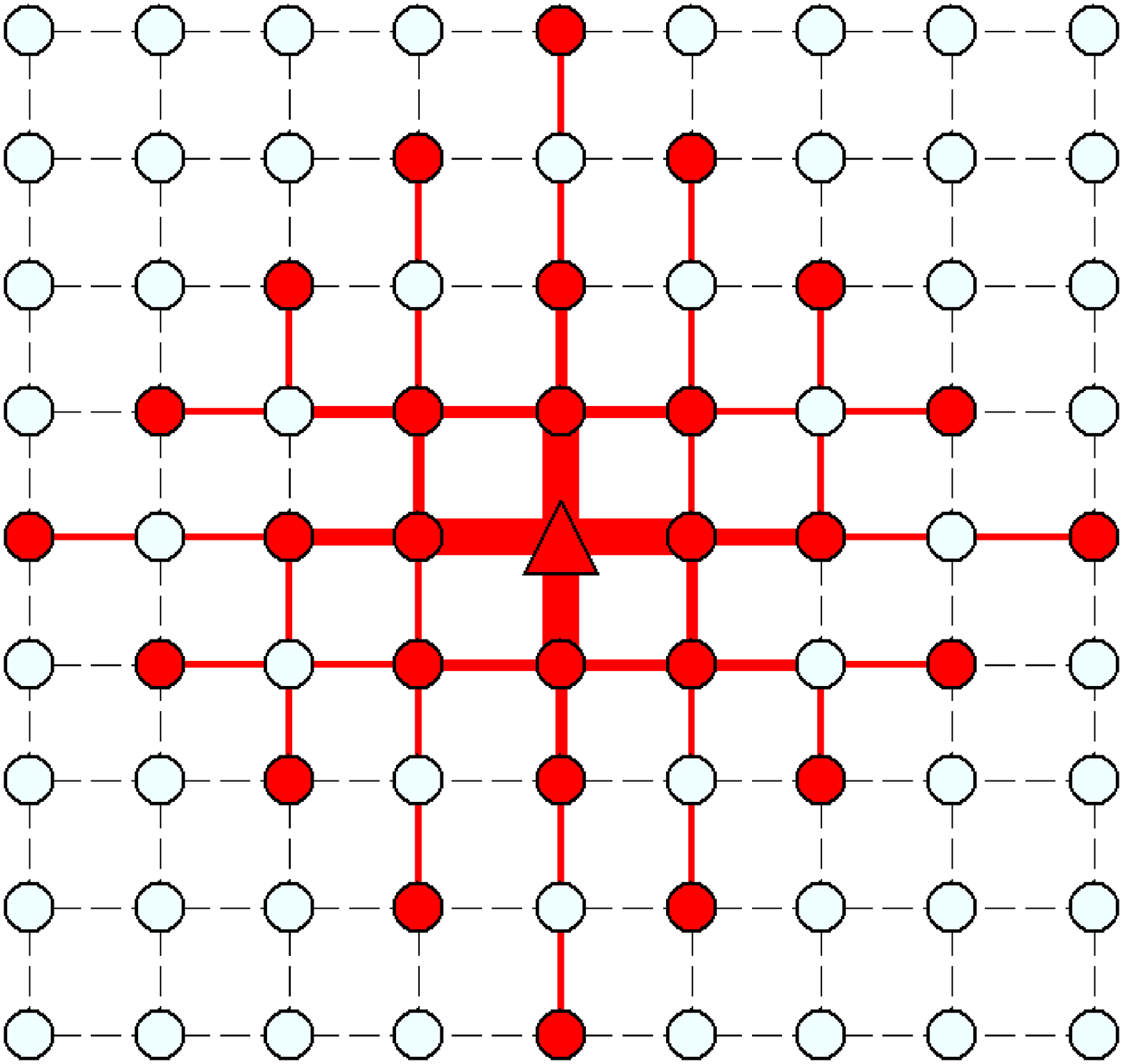, width=0.6\linewidth}}
\vspace{0.2cm}
\leftline{\hspace{1.2cm}(b)}
\vspace{-0.3cm}
\centerline{\epsfig{figure=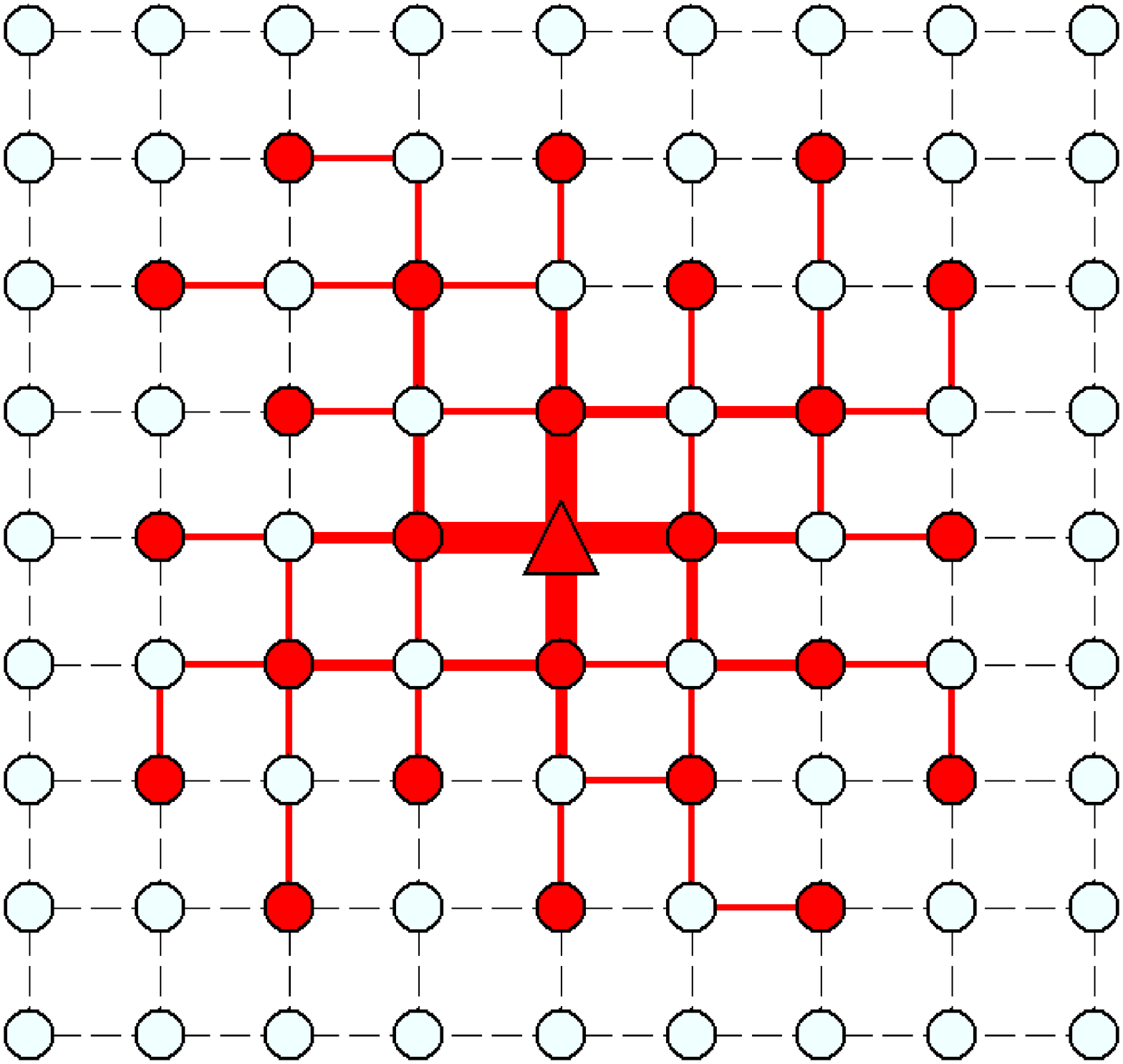, width=0.6\linewidth}}
\caption{
(Color online) The optimized node configuration on a square lattice with 81 nodes, (a)  $J = 0.3$ and $U = 20$, (b) $J = 5$ and $U=0$. The triangle, red filled nodes and empty nodes correspond to the terminal, active and idle nodes respectively. Dashed edges are idle, and the thickness of the solid edges is proportional to $|I|$, the flow on individual edges. }
\label{fig_ex}
\end{figure}

To further illustrate the analogy in this paper, we will consider the Hamiltonian on the square lattice and random regular graphs. Spin models with competing interactions on the square and other regular lattices have been studied in the Axial Next Nearest Neighbor Ising (ANNNI) Model~\cite{selke88}, and the Next Nearest Neighbor Ising model (NNNI)~\cite{domb51}. When the ratio of the nearest to next nearest couplings (both antiferromagnetic) changes, there are phase changes in the periodicity of the modulated configuration, and may even lead to the absence of long range order at some special coupling ratio. At non-zero temperature, further modulation patterns appear, and there are metastable states~\cite{stephenson70}. The periodicity of the modulations has infinite steps when the coupling ratio changes, exhibiting the so-called Devil's staircase behavior~\cite{bak80}. All these point to the existence of many metastable states produced by frustration. Progress on the analysis of many ANNNI and NNNI models was made by approximating the local structures to be tree-like~\cite{vannimenus81}. They reproduce many thermodynamic features of the finite dimensional NNNI models on random graphs including the special ratios~\cite{inawashiro83, ganikhodjaev03}.

The new observation in our model is the coexistence of active, idle and antiferromagnetic domains in a single network. We show in \fig{fig_ex}(a) an example of the ground state of $H$ on a square lattice. One can identify that (i) the emergence of a core of active nodes is due to the presence of an external field, (ii) the alternating active-idle outer layer is due to the antiferromagnetic interaction, and (ii) the four idle corner regions are due to the presence of supply cost.

We note that there are some fundamental differences between the present model and the conventional facility location problem, such that algorithms of the latter cannot be readily applied to solve the problem. Specifically, the presence of the antiferromagnetic term to encourage the spread of facilities have significantly increased the extent of frustration in the system. Besides, the quadratic transportation cost which discourages active nodes from sharing their path is a major feature of the present formulation. Among the various versions of conventional facility location problems, the version comparable to our model would have the objective of minimizing the population deprived of facilities \cite{pirkul98,facility,moore82}. However, these problems paid little attention to whether the deprived population is clustered topographically, or to the cost incurred by congested transportation paths. These differences render the application of existing algorithms on the present model difficult and makes the algorithm derived in Section III essential. Nevertheless, we have compared the performance of our algorithm with the IBM CPLEX solver~\cite{cplex} and the results are discussed in Section III C.

\section{The cavity equations}

Based on the tree-like picture, we will obtain the optimal state of the present model by employing the cavity approach developed in the field of spin glasses and disordered systems~\cite{mezard87, nishimori01}. The cavity approach assumes that only large loops exist in a network such that neighbors of a node become independent. In regular lattices, the cavity approach is effectively only approximate. Nevertheless, we found that the derived algorithm yields accurate results in networks with short loops, including regular lattices. We will see in Section~\ref{sec_sq} that the algorithmic solution is identical to the known optimal solution in simple cases. On random regular graphs,  the absence of short loops renders the application of cavity approach more justified. In Section~\ref{sec_rrg} we will find more evidence of frustration, and a transition from replica symmetric to replica symmetry-breaking phase.

To obtain various macroscopic properties, we first denote $E_{i\to l}(s_i, I_{il})$ as the optimized energy of the tree terminated at the edge from node $i$ to $l$,  where the  state of node $i$ and the current to node $l$ are $s_i$ and $I_{il}$ respectively. This energy includes the supply cost $\phi(I_{ij})$ but exclude the coupling term $Js_is_l$. One can write down a recursive relation relating $E_{i\to l}(s_i, I_{il})$ to the energy $E_{j\to i}(s_j, I_{ji})$ from neighbor $j$ to node $i$, i.e.
\begin{align}
\label{eq_cavity}
&E_{i\to l}(s_i, I_{il}) = \phi(\flow_{il}) +
\\
&\min_{\{\{s_j,I_{ji}\}|f_i=I_{il}\!\}}
\hspace{-0.18cm}
\left[\!\sum_{j\in\!\cN_i\backslash l}\!Js_is_j
\!+\!U\frac{1\!-\!s_i}{2}
\!+\hspace{-0.3cm}\sum_{j\in \cN_i\backslash l}\!E_{j\to i}(s_j,I_{ji})\!\right]\!,
\nonumber
\end{align}
where $f_i=\sum_{j\in \cN_i\backslash l}I_{ji}-\frac{1+s_i}{2}$. We have assumed that the minimization on the right hand side of \req{eq_cavity} does not depend on the states $s_k$ of the next nearest neighbors $k\in \cN_j\backslash i$. As we shall see, it turns out that the assumption yields good results. For the terminal node $T$, we assume $s_T = -1$ since it does not serve as an active node. The optimized energy function $E_{T\to l}(s_T, I_{Tl})$ from the terminal to its neighbor $l$ is given by
\begin{align}
E_{T\to l}(s_T, I_{Tl}) = \phi(\flow_{Tl}) + C(1+s_T),
\end{align}
for some arbitrary large constant $C > 0$ such that $s_T = -1$ always has a lower energy compared to $s_T = 1$. 

Since $E_{i\to l}(s_i, I_{il})$ is an extensive quantity, we define its intensive counterpart by $E^V_{i\to l}(s_i, I_{il})=E_{i\to l}(s_i, I_{il})-\min_{\{s'_i, I'_{il}\}} [E_{i\to l}(s'_i, I'_{il})]$, and further denote the right hand side of \req{eq_cavity} as
\begin{align}
\label{eq_cM}
&\cM\left[s, I; \{E_{j\to i}\}_{j\in \cS}\right] \equiv  \phi(\flow)+
\\
&
\min_{\{\{s_j, I_{ji}\}|f_i=I\}}\left[\sum_{j\in \cS}Jss_j+U\frac{1-s}{2}
+\sum_{j\in \cS}E_{j\to i}(s_j, I_{ji})\right]\!,
\nonumber
\end{align}
where $\cS$ denotes a set of nodes. The recursive equation of $E^V_{i\to l}(s_i, I_{il})$ in \req{eq_cavity} can be written as
\begin{align}
\label{eq_EV}
E^V_{i\to l}(s_i, I_{il}) &= \cM\left[s_i, I_{il}; \{E^V_{j\to i}\}_{j\in \cN_i\backslash l}\right] 
\nonumber\\
&\hspace{0.5cm}- \min_{s, I}\left\{\cM\left[s, I; \{E^V_{j\to i}\}_{j\in \cN_i\backslash l}\right]\right\}.
\end{align}
One can obtain a stable functional distribution $P[E^V(s, I)]$ of $E^V(s, I)$ by the following self-consistent equation
\begin{align}
P[E^V(s, I)]\!&=\!\sum_{k=1}^{\infty}\frac{k\rho(k)}{\avg{k}}\int\prod_{\substack{j=1\\s_j, I_j}}^{k} d E^V_{j}(s_j, I_{j}) P[E^V_{j}(s_j, I_{j})]
\nonumber\\
&\times\prod_{s, I}\delta\bigg(E^V(s, I) - \cM\left[s, I; \{E^V_{j}\}_{j=1,\dots,k}\right] 
\nonumber\\
&\hspace{0.5cm}+\min_{s',I'}\left\{\cM\left[s', I'; \{E^V_{j}\}_{j=1,\dots,k}\right]\right\}\bigg),
\end{align}
where $k\rho(k)/\avg{k}$ is the so-called \emph{excess degree distribution}, which describes the probability of finding an edge connected to a node of degree $k$. The above equation can be solved, for instance, by population dynamics. By averaging over $P[E^V(s, I)]$, the optimized energy can be computed by $\avg{E} \!=\! (\avg{E_{\rm node}}\!-\!\frac{\avg{k}}{2}\avg{E_{\rm link}})$ following the conventional cavity approach~\cite{mezard02}, where $\avg{E_{\rm node}}$ and $\avg{E_{\rm link}}$ are given by
\begin{align}
\label{eq_enode}
&\avg{E_{\rm node}} = \Big\langle \min_{s}\left\{\cM\left[s, 0; \{E^V_{j\to i}\}_{j\in \cN_i}\right]\right\}\Big\rangle_{\{E^V_{j\to i}\}_{j\in \cN_i}},
\\
\label{eq_elink}
&\avg{E_{\rm link}} = 
\\
&\Big\langle\!\min_{s_1, s_2, \flow}\!\left\{E^V_{j_1}(s_1,\flow)\!+\!E^V_{j_2}(s_2,-\flow)\!-\!\phi(\flow)\!+\!Js_1s_2\right\}\!\Big\rangle_{E^V_{j_1},E^V_{j_2}}.
\nonumber
\end{align}

\section{Algorithms and local expansion of messages}
\label{sec_algorithm}

The recursion of $E^V_{j\to i}$ in \req{eq_EV} gives rise to a message passing algorithm. To optimize the total cost, one randomly chooses a node $i$ and gathers the functional messages $E^V_{j\to i}$ from all neighbors $j$ of $i$. We then randomly choose one of the neighbors to be the ancestor $l$, and update the message $E^V_{i\to l}$ according to \req{eq_EV}. The procedure continues until the messages converge on all edges in both directions, i.e. $E^V$ does not change in consecutive updates for a sufficiently large number of steps. Using equations resembling \req{eq_elink}, one can compute the optimal flow $I_{il}^*$ on the edge from $i$ to $l$ by
\begin{align}
\label{eq_opti}
\flow_{il}^* = \argmin_{s_i, s_l, \flow}\left\{E^V_{i\to l}(s_i, \flow)+E^V_{l\to i}(s_l, -\flow)-\phi(\flow)+Js_is_l\right\}.
\end{align}
The optimal state $s_i^*$ of a node $i$ is given by 
\begin{align}
\label{eq_opts}
s_i^* = \argmin_{s}\left\{\cM\left[s, 0; \{E^V_{j\to i}\}_{j\in \cN_i}\right]\right\}.
\end{align}
Alternatively, one can also compute $s_i^*$ using \req{eq_conserve} and $\{\flow_{ji}^*\}$ obtained from \req{eq_opti}, such that
\begin{align}
\label{eq_sis}
s_i^* = -1+2\sum_{j\in\cN_i}I^*_{ji}.
\end{align}
Once the optimal flow on all the edges is obtained, the communication path from individual active nodes to the terminal can be identified by tracing from the active node a path to the terminal through edges of non-zero flow un-occupied by other communications.

Although the algorithm is easily formulated by Eqs. (\ref{eq_EV}), (\ref{eq_opts}) and (\ref{eq_opti}), it is difficult to compute since (i) the algorithm involves functional messages with indefinite domain, and (ii) the recursion \req{eq_EV} involves constrained optimization on integer domains.

\subsection{Local expansion of messages}

To cope with these difficulties, we employ an approach introduced in~\cite{wong06} to simplify the message passing algorithm. Since the derivation is rather involved, we refer readers who would like to skip the derivation to the summary of the algorithm below \req{eq_sigmajs}.

To simplify the present message-passing algorithm, we first note that in~\cite{wong06}, an optimization algorithm is derived for a resource allocation problem involving functional messages. Unlike the present case, the messages in~\cite{wong06} have a continuous domain, which renders their exact computation infeasible except for some specific form of cost function. Nevertheless, instead of a complete functional message, only the values on a small domain are necessary to verify the optimality of the converged state. This allows Wong and Saad~\cite{wong06} to parametrize the functional messages by only the first two coefficients in its Taylor's series expansion around a \emph{working point}, which is given by the ancestor. The working points are continuously updated by the ancestors until they coincide with the optimal state. Here we employ a similar idea of local expansion to effectively parametrize the functional messages on integer domains. The same approach has been adopted in routing optimization in~\cite{yeung13c}. To locally expand the messages, we define
\begin{align}
\label{eq_hil}
\hilm &= E^V_{i\to l}(1, I^*_{il}+\sigma) -  E^V_{i\to l}(-1, I^*_{il}),
\\
\label{eq_gil}
\gilm &= E^V_{i\to l}(-1, I^*_{il}+\sigma) -  E^V_{i\to l}(-1, I^*_{il}),
\end{align}
where $I^*_{il}$ is the working point of flow given by node $l$ for node $i$, and $-M\le\sigma\le M$ is an integer. The parameter $M$ determines the range of flows around the working point to be parametrized, the smaller the the value of $M$, the fewer  variables used to parametrize the functional messages $E^V_{i\to l}(s, I)$. When $M\to\infty$, the complete profile of $E^V_{i\to l}(s, I)$ is recovered. We observed that the algorithmic performance does not improve beyond $M=2$, such that $M=2$ is already sufficient for the systems studied. We have chosen $E^V_{i\to l}(-1, I^*_{il})$ to be a common reference point in both Eqs.~(\ref{eq_hil}) and (\ref{eq_gil}), while other choices would yield identical physical results. In this case, $h_{i\to l}^{(0)}=E^V_{i\to l}(1, I^*_{il})-E^V_{i\to l}(-1, I^*_{il})$ and $g_{i\to l}^{(0)}=0$. The $h_{i\to l}$ and $g_{i\to l}$ variables can be regarded as the cavity fields at zero temperature. 

We then continue to derive a recursion relation involving only $h$ and $g$. We first denote the vectors $\left(\hil^{(-M)}, \hil^{(-M+1)}, \dots, \hil^{(M)}\right)$ and  $\left(\gil^{(-M)}, \gil^{(-M+1)}, \dots, \gil^{(M)}\right)$ by $\vhil$ and $\vgil$, respectively. From Eqs.~(\ref{eq_EV}) and (\ref{eq_hil}), one can express $\hilm$ as
\begin{align}
\label{eq_h}
\hilm &= \cM\left[1, I^*_{il}\!+\!\sigma; \{E^V_{j\to i}\}_{j\in \cN_i\backslash l}\right] 
\nonumber\\
&\hspace{2cm}
- \cM\left[-1, I^*_{il}; \{E^V_{j\to i}\}_{j\in \cN_i\backslash l}\right] 
\nonumber\\
&\approx
\phi(I^*_{il}+\sigma) - \phi(\Isil) + \cQ\left[1, \sigma, \Isil; \hjinl, \gjinl, \Isjinl\right] 
\nonumber\\
&\hspace{2cm}
- \cQ\left[-1, 0, \Isil; \hjinl, \gjinl, \Isjinl\right] 
\end{align}
where $\hjinl=\{\vhji| j\in \cN_i\backslash l\}$ and $\gjinl=\{\vgji| j\in \cN_i\backslash l\}$. The set $\Isjinl=\{I_{ji}^*| j\in \cN_i\backslash l\}$ contains the working points where the messages $\hjinl$ and $\gjinl$ were computed. The functional $\cQ$ is defined as
\begin{align}
\label{eq_cQ}
&\cQ\left[s, \sigma, I^*; \hjinl, \gjinl, \Isjinl\right] = \min_{\{\{s_j, \sigma_j\}|f_i=I^*+\sigma\}}\left[\sum_{j\in \cN_i\backslash l}Jss_j
\right.
\nonumber\\
&\left.
+U\frac{1-s}{2}+\sum_{j\in \cN_i\backslash l}\left(\hji^{(\sigma_j)}\delta_{s_j,1}+\gji^{(\sigma_j)}\delta_{s_j,-1}\right)\right]\!,
\end{align}
where $f_i=\sum_{j\in \cN_i\backslash l}(I^*_{ji}+\sigma_j)-\frac{1+s}{2}$. The approximation sign in \req{eq_h} comes from the restricted range of $-M\le\sigma_j\le M$ when $M$ is finite, such that the minimization in $\cQ$ in \req{eq_cQ} is only an approximation of $\cM$ in \req{eq_cM}. In other words, \req{eq_h} becomes exact only when $M\to\infty$. Similarly, from Eqs.~(\ref{eq_EV}) and (\ref{eq_gil}), one can express $\gilm$ as
\begin{align}
\label{eq_g}
\gilm &\approx
\phi(\Isil+\sigma) - \phi(\Isil) + \cQ\left[-1, \sigma, \Isil; \hjinl, \gjinl, \Isjinl\right] 
\nonumber\\
&\hspace{2cm}
- \cQ\left[-1, 0, \Isil; \hjinl, \gjinl, \Isjinl\right].
\end{align}

In addition to the update of messages $\vhil$ and $\vgil$ from node $i$ to ancestor $l$, node $i$ has to update the working points for its descendents $j$ at which new messages $\vhji$ and $\vgji$ will be computed in the next round. These working points are updated by 
\begin{align}
\label{eq_Isji}
\Isji +\sigma_j^* \to \Isji
\end{align}
where
\begin{align}
\label{eq_sigmajs}
&\sigma_j^*=\argmin_{\{s_i, \{s_j, \sigma_j\}|f_i=I^*_{il}\}}\left[\sum_{j\in \cN_i\backslash l}Js_is_j
\right.
\nonumber\\
&\left.
+U\frac{1-s_i}{2}+\sum_{j\in \cN_i\backslash l}\left(\hji^{(\sigma_j)}\delta_{s_j,1}+\gji^{(\sigma_j)}\delta_{s_j,-1}\right)\right]\!.
\end{align}

In summary, the algorithm proceeds as follows:
\begin{enumerate}
\item
Randomly draw a node $i$ and choose at random one of its neighbors to be the ancestor $l$.
\item
Obtain the updated working point $\Isil$ from node $l$ and gather $\vhji$, $\vgji$ and the corresponding working point $\Isji$ at which $\vhji$ and $\vgji$ were computed from all the neighbors of $i$ other than $l$.
\item
Compute $\vhil$ and $\vgil$ by Eqs.~(\ref{eq_h}) and (\ref{eq_g}), respectively. Update the working points $\Isji$ by Eqs.~(\ref{eq_Isji}) and (\ref{eq_sigmajs}) for all neighbors $j$ except $l$.
\item
Repeat steps 1 to 3 until all messages and working points converge, i.e., do not change for a sufficiently long time.
\item
The optimal flows on the edges  are given by the converged working points. The optimal state of node $i$ is given by  \req{eq_sis}. The communication path between individual active nodes and the terminal is identified by tracing from each node $i$ through edges with non-zero flow un-occupied by  the communication paths of other active nodes.
\end{enumerate}

\subsection{Random biases}
\label{sec_bias}

As suggested by previous models of communication networks~\cite{yeung12, yeung13, yeung13c}, the presence of degenerate optimized states may hinder algorithm convergence. One can adopt the same approach to break degeneracy by assigning to each edge $(ij)$ a small random bias $\epsilon_{(ij)}$ and optimize a cost $\phi(\flow_{ij})+|\flow_{ij}|\epsilon_{(ij)}$, which helps the algorithm converge to one of the optimized states.

In the parameter regimes where the algorithm fails to converge even with the set of small $\epsilon_{(ij)}$ implemented, we found that the following procedures may improve convergence:
\begin{enumerate}
\item
Start with a set of large $\epsilon_{(ij)}$ such that the algorithm converges. Obtain a converged state.
\item
Decrease the bias by $\epsilon_{(ij)}\to f\epsilon_{(ij)}$ for some fraction $f<1$.
\item
Use the previous converged state as the initial condition to obtain a new convergent state with the new set of biases.
\item
Repeat step 1 - 3 to decrease the value of $\epsilon_{(ij)}$ until the algorithm fails to converge or $\epsilon_{(ij)}$ is smaller than a threshold. The converged solution obtained by the set of smallest $\epsilon_{(ij)}$ is considered as the optimized configuration.
\end{enumerate}
By repeating step 1 to 3, we found that the energy of the converged state is decreasing with decreasing $\epsilon_{(ij)}$.  In regimes where the algorithm converges without the above procedures, in most cases we obtain an identical algorithmic solution  with the above process. We thus expect that an optimal or close-to-optimal state is obtained in the regime where the algorithm ceases to converge with small $\epsilon_{ij}$. Physically, these procedures are equivalent to the case when one applies a strong external field to facilitate the convergence to an anticipated state, and slowly decrease the field such that the system is attracted to one of the local minima nearby.

\subsection{Comparison with CPLEX}

We have compared the performance of our algorithm with the IBM CPLEX solver~\cite{cplex}, which is capable of solving integer and quadratic programming problems. We found that in parameter regimes with intermediate idle penalties and non-zero anti-ferromagnetic couplings, the computational complexity of the solver scales up rapidly, or even exponentially increases with system size. It leads to a long running time even for relatively small size of $O(10^2)$. On the other hand, our algorithm does not converge on all instances in the same regime, but it does show a quadratic complexity with system size by averaging over the converged realizations. For those non-converged instances which cannot be solved by the approach described in the previous subsection, techniques such as decimation can be applied to find a solution~\cite{mezard02}. In general, our algorithm compares favorably with integer quadratic programming techniques in terms of computational complexity.

\section{Results}

\subsection{Square Lattice}
\label{sec_sq}

We first examine the optimized location of active nodes on a square lattice where the terminal is located at the center.  We note that the algorithm derived in Section~\ref{sec_algorithm} is based on the cavity approach which assumes the absence of short loops, and is only a crude approximation on square lattices. Nevertheless, the algorithm converges in a majority of the parameter regime. For regimes where the algorithm ceases to converge, we found that the procedures suggested in Section~\ref{sec_bias} facilitate the algorithmic convergence.

\begin{figure}
\leftline{\hspace{0.7cm}(a)\hspace{3.15cm}(d)}
\vspace{-0.2cm}
\centerline{\epsfig{figure=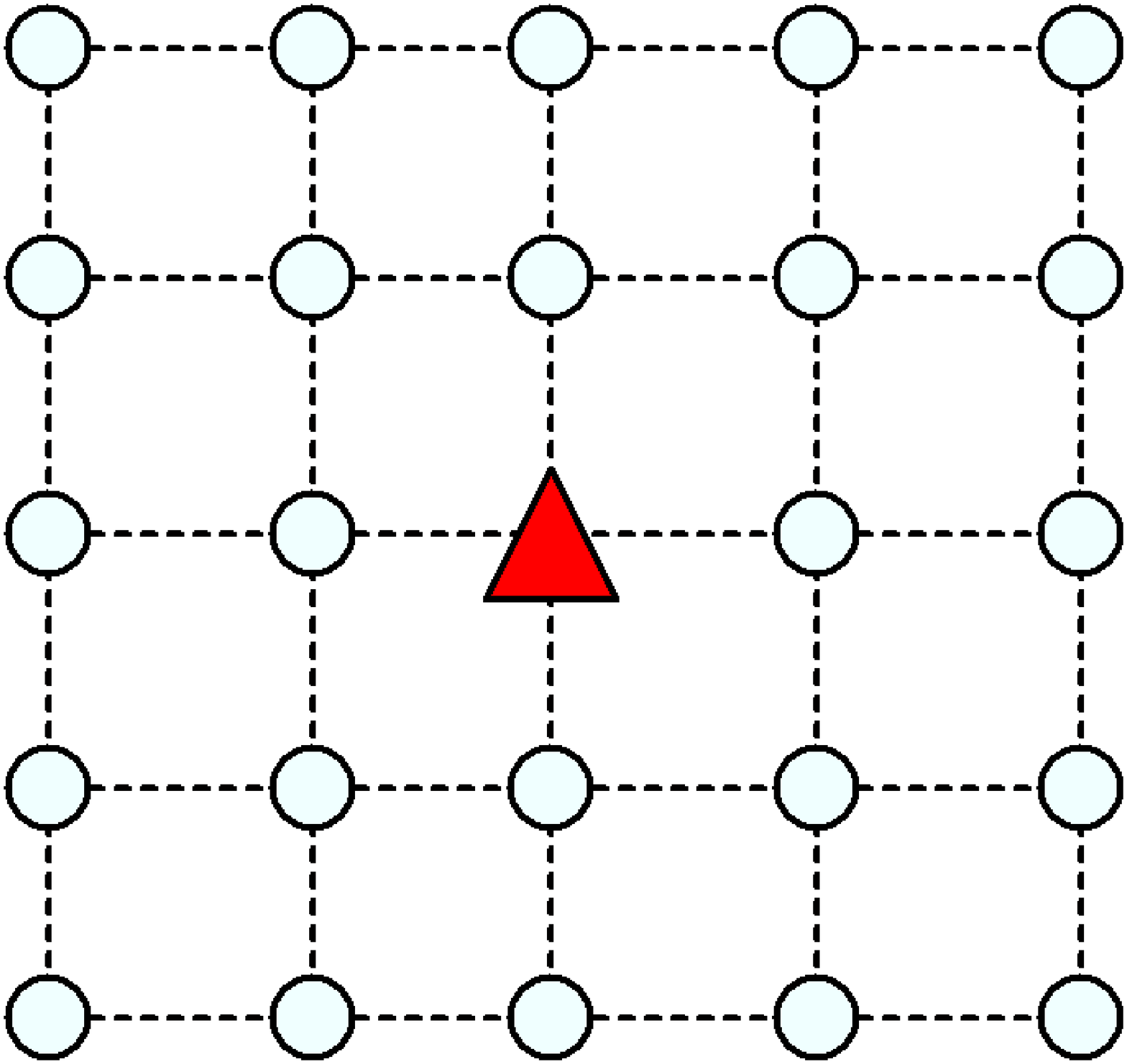, width=0.3\linewidth}\hspace{1cm}\epsfig{figure=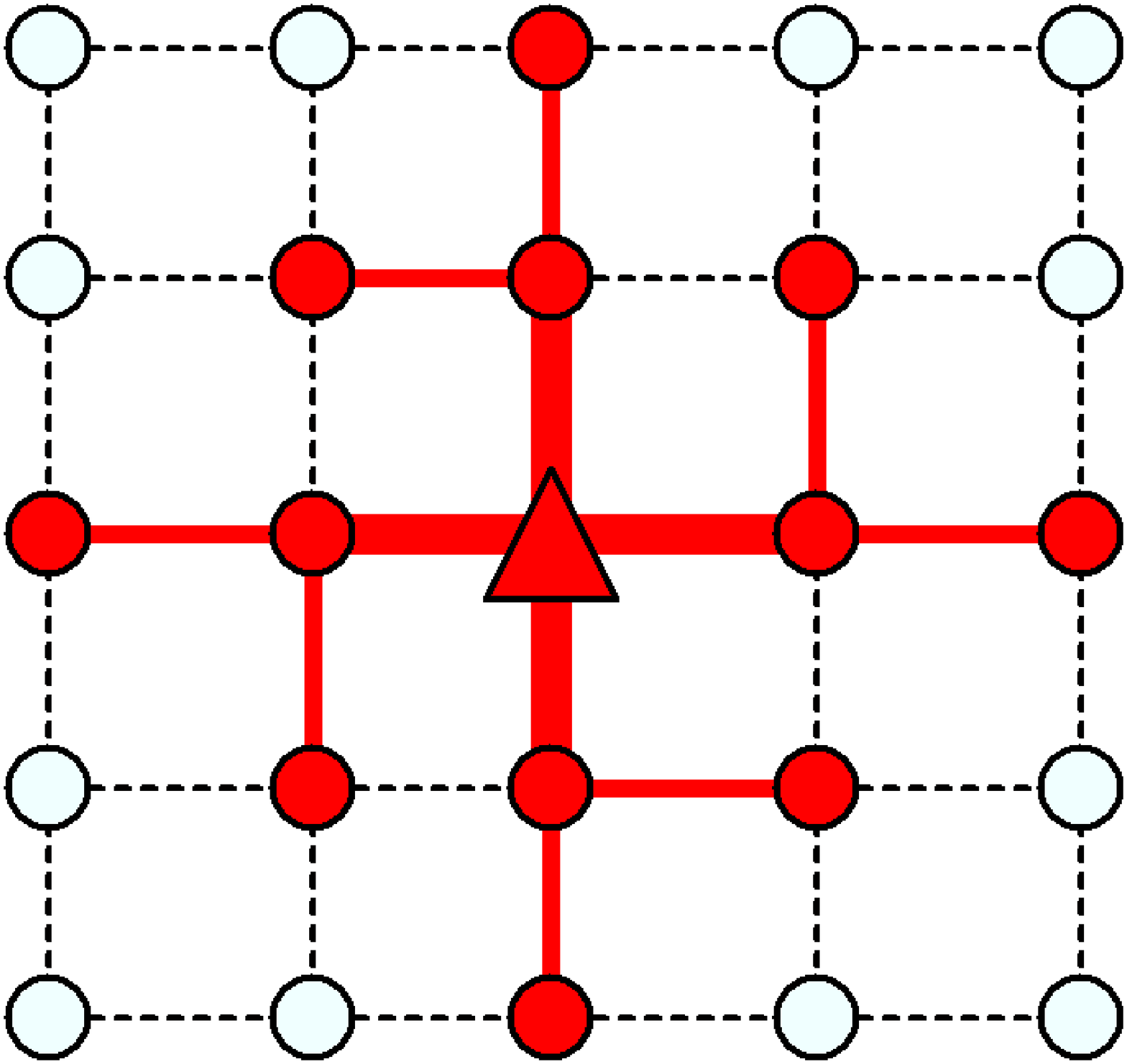, width=0.3\linewidth}}
\vspace{0.2cm}
\leftline{\hspace{0.7cm}(b)\hspace{3.15cm}(e)}
\vspace{-0.2cm}
\centerline{\epsfig{figure=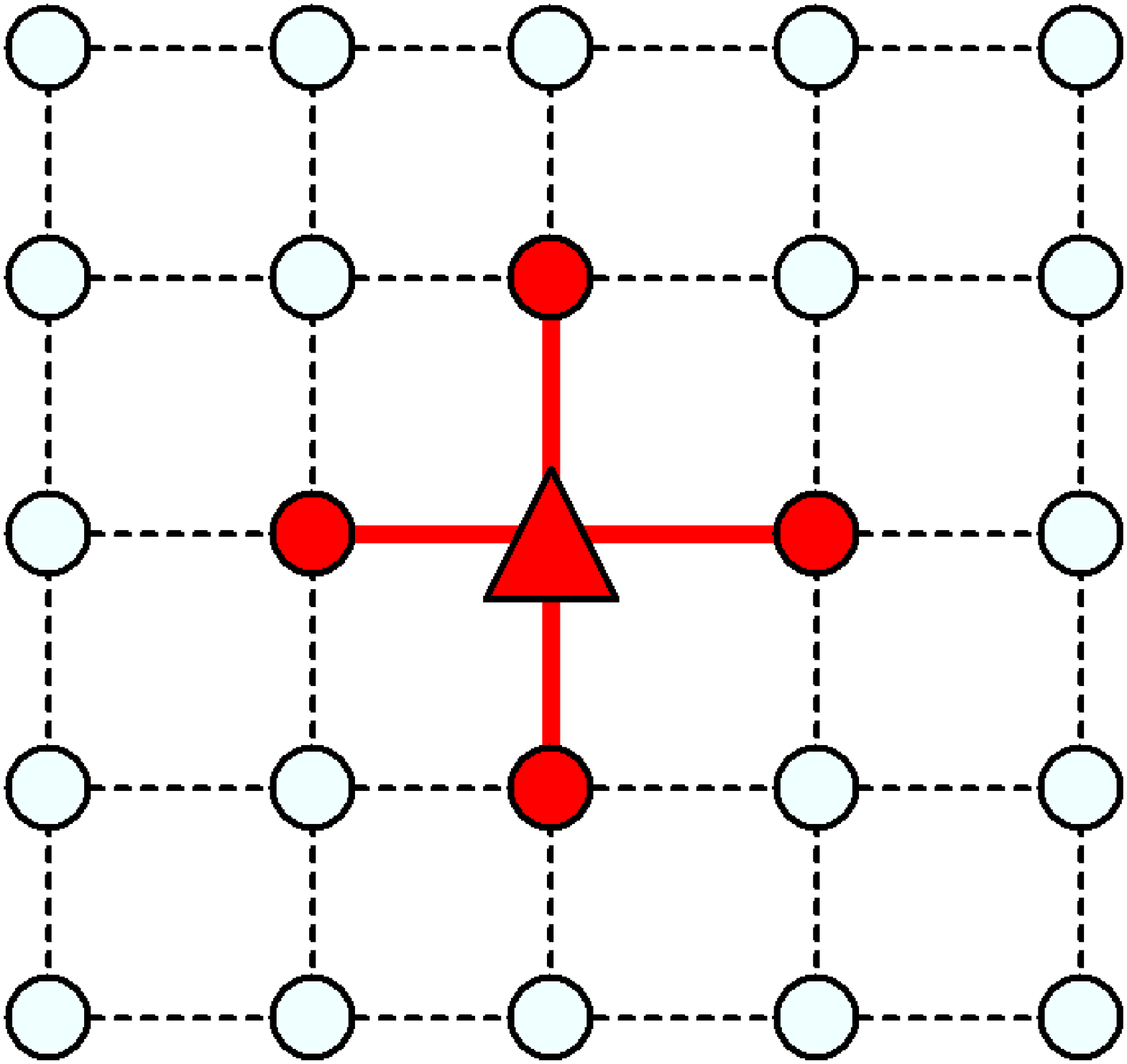, width=0.3\linewidth}\hspace{1cm}\epsfig{figure=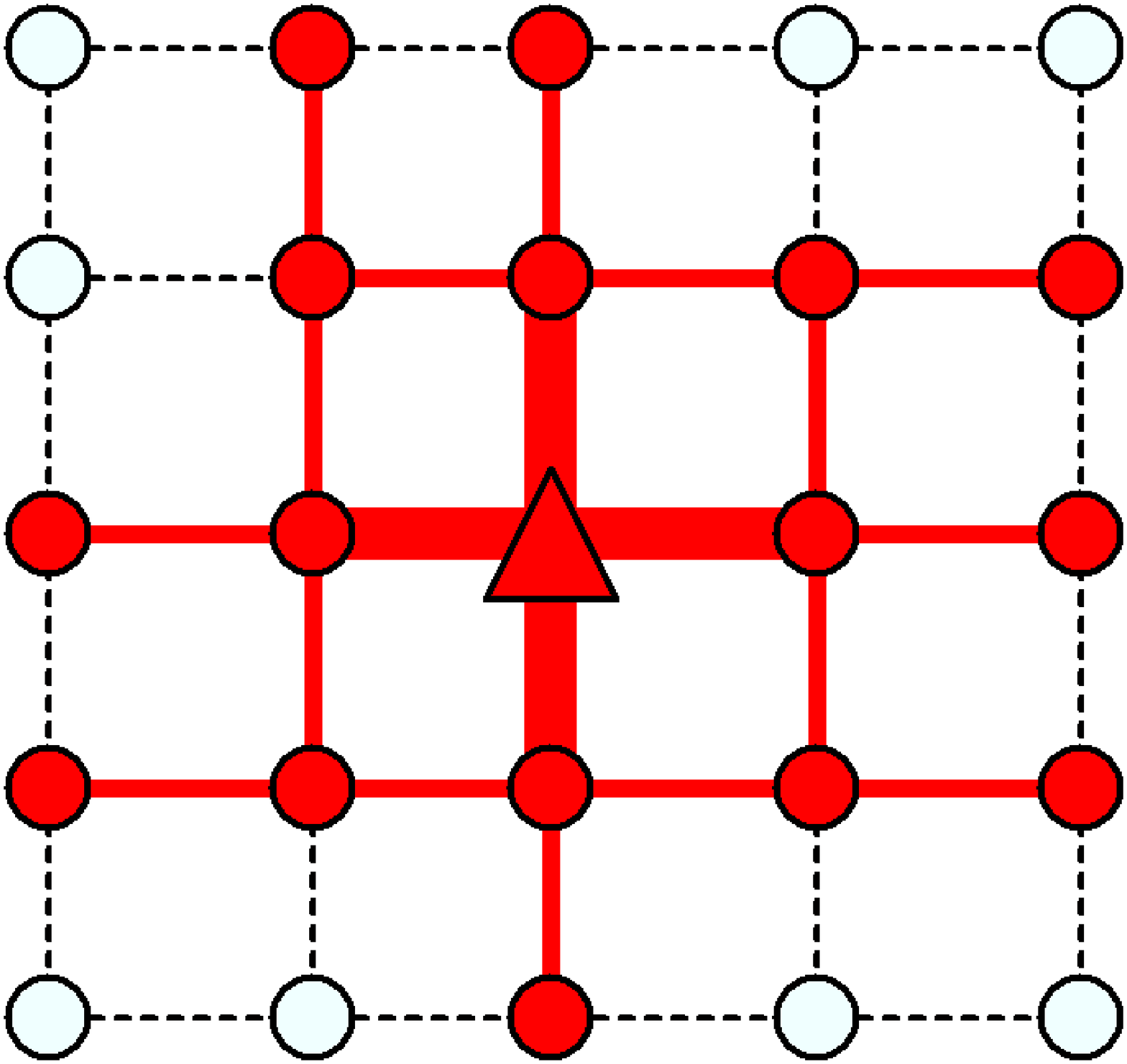, width=0.3\linewidth}}
\vspace{0.2cm}
\leftline{\hspace{0.7cm}(c)\hspace{3.15cm}(f)}
\vspace{-0.25cm}
\centerline{\epsfig{figure=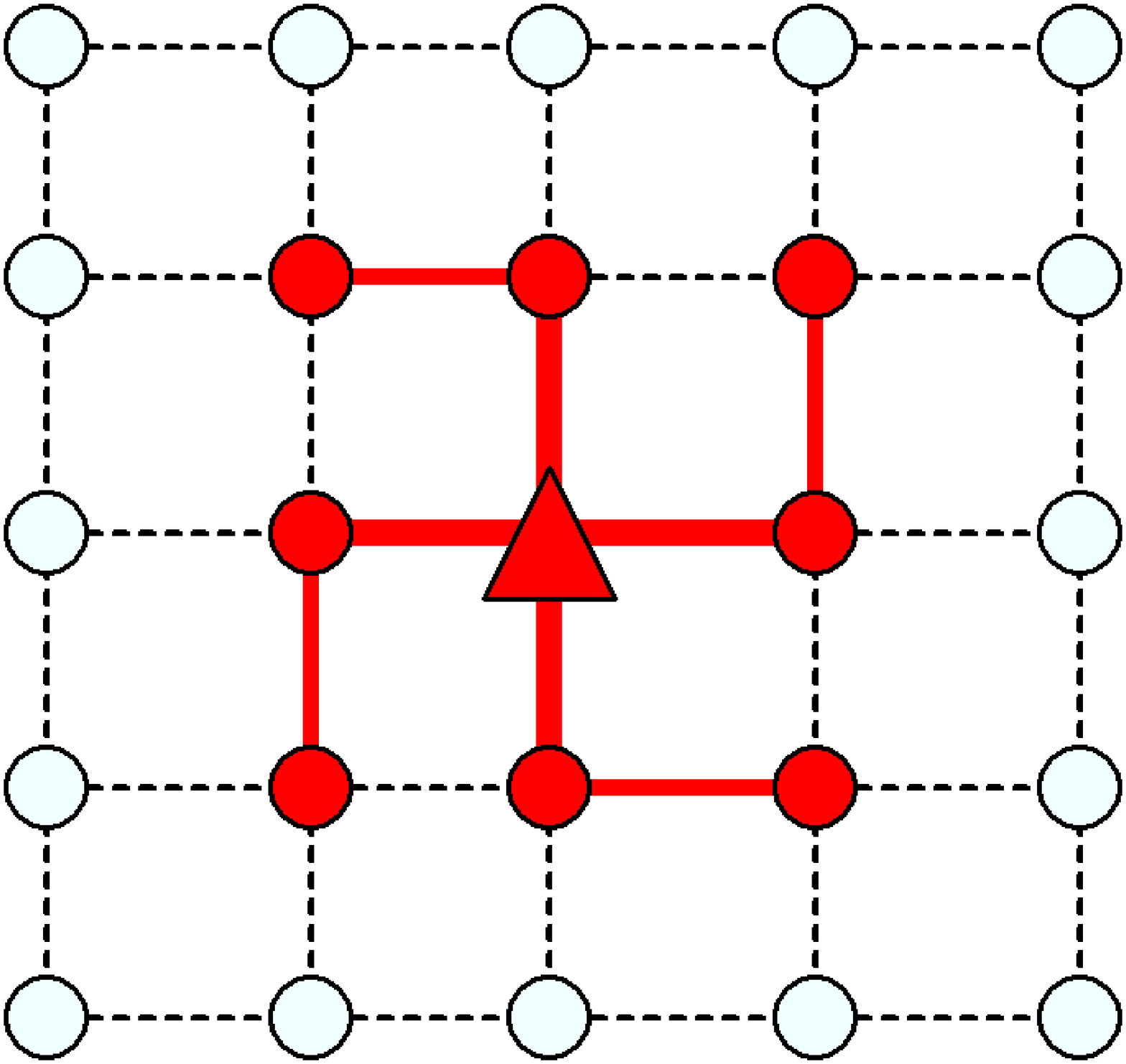, width=0.3\linewidth}\hspace{1cm}\epsfig{figure=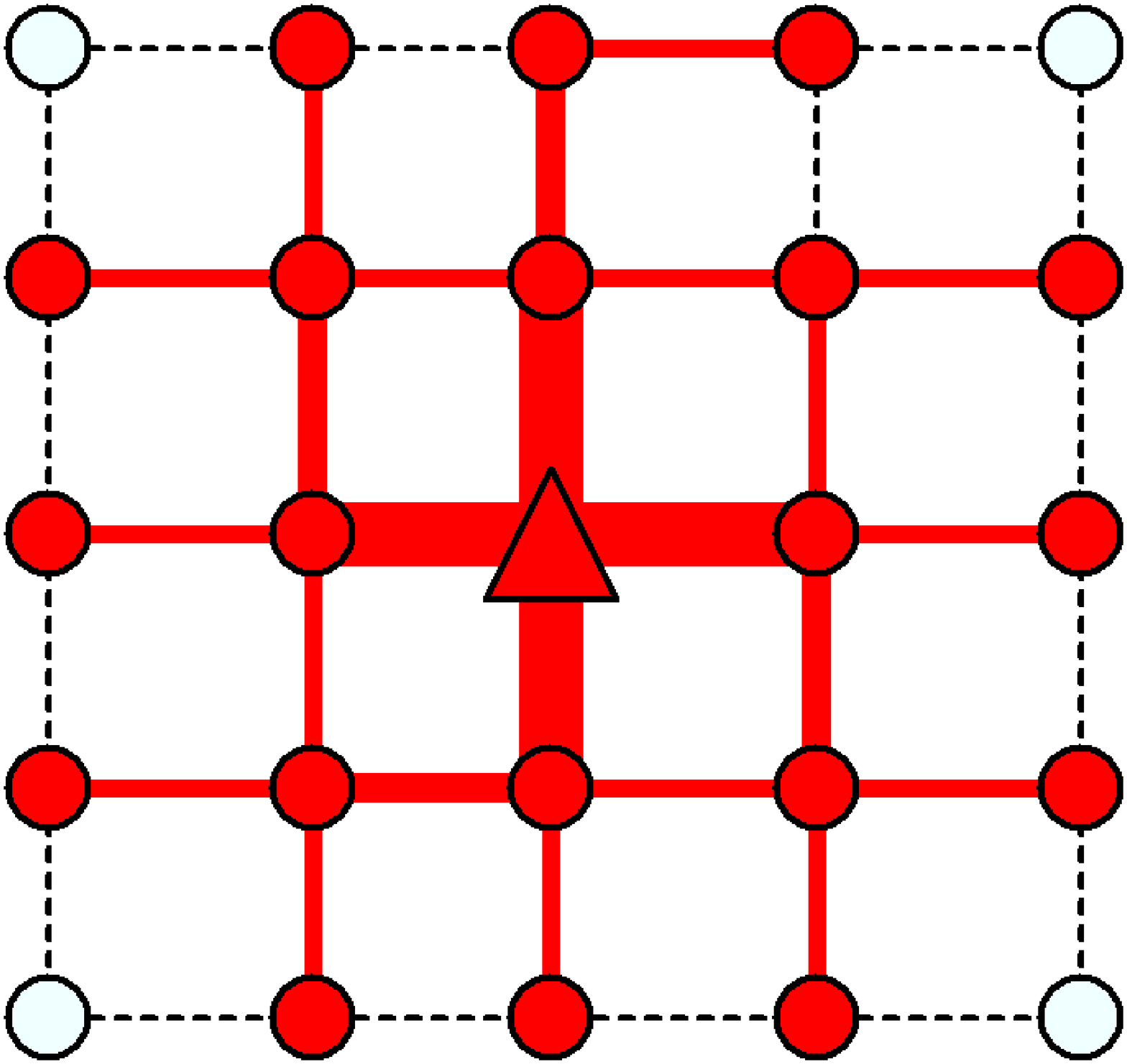, width=0.3\linewidth}}
\caption{
 (Color online) The optimized  node configuration on a square lattice of $N = 25$ with $J = 0$ and (a) $U = 0.5$, (b) $U = 2$, (c) $U = 5$, (d) $U = 7$, (e) $U = 10$ and (f) $U = 14$. Other  configurations with energy identical to (c) and (e) exist.}
\label{fig_simple}
\end{figure}

To verify the algorithmic results on square lattices, we examine a simple system of size $N = 25$ with $J = 0$. As shown in \fig{fig_simple}(a)-(f), a core of active nodes grows around the center when $U$ increases. In this case, one can analytically compute the number of active nodes $N_a$ as a function of $U$. For instance, when $U$ increases gradually from $U=0$, we expect the system to change at a threshold value from the complete idle state in \fig{fig_simple} (a), to the state where the four neighbors of the terminal become active, as shown in \fig{fig_simple}(b). To compute the threshold value, we consider the change in energy of the configuration from \fig{fig_simple}(a) to (b)   given by $[(N -5)U +4]-(N - 1)U$,   a negative energy change at $U = 1$ signals the stability of \fig{fig_simple}(b) over (a) such that $N_a$ increases from 0 to 4. We remark that there is no intermediate optimized configuration with $1\le N_a\le 3$ since the system  has a 4-fold symmetry; if an active node is energetically favorable in one direction, then other active nodes are  also favorable in the other three directions.

Similarly, one can obtain other threshold values of $U$ by considering the energetic stability of consecutive configurations in \fig{fig_simple}. Degenerate states, i.e. states with the same optimized energy but different configuration, do exist for \fig{fig_simple}(c) and (e). Nevertheless, their existence has no influence on the threshold values.

\begin{figure}
\centerline{\epsfig{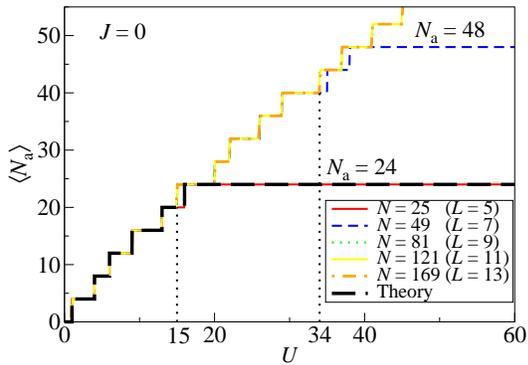}}
\caption{
 (Color online) The number of active nodes $\avg{\Non}$ as a function of idle-penalty $U$ for $N = 25, 49, 81, 121$ and 169, obtained by the algorithm derived by the cavity approach. Results are averaged over 10 realizations with different sets of random biases. The analytical results for $N=25$ are obtained by identifying the active node configuration with lowest energy (shown in \fig{fig_simple}) at specific values of $U$.}
\label{fig_step}
\end{figure}

	In \fig{fig_step}, we show the excellent agreement between the results from the message-passing algorithm and the above analytical $N_a$. The agreement suggests that, although  the cavity approach is an  approximation on square lattices, the derived algorithm leads to the optimal state at $J = 0$. We thus expect that an optimal state, or at worst a close-to-optimal state, is obtained by the algorithm when it converges. Other than the square lattice of size $N = 25$, the analytical $N_a$ also agrees well with algorithmic results in larger systems, suggesting identical active node configurations in the range $0 < U < 15$ for systems of all sizes. Differences between $N = 25$ and larger lattices emerge beyond $U = 15$, since larger systems can attain a lower energy state with a diamond-shaped active core instead of  an active square with size $N=25$.  Similar differences are observed between $N = 49$ and larger lattices beyond $U = 34$. These differences are  due to boundary effects of the smaller lattices.

\subsubsection{The various phases}

\begin{figure}
\centerline{\epsfig{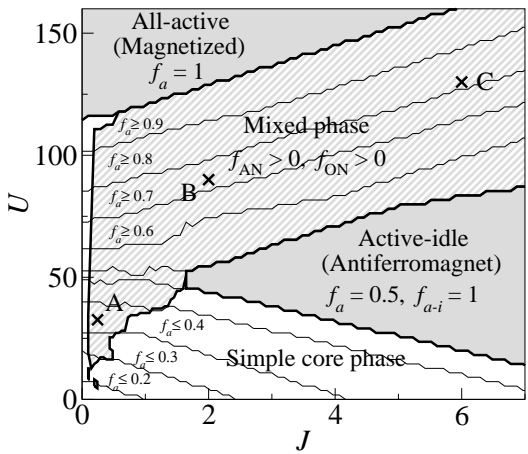}}
\caption{
Phase diagram and the fraction of active nodes as a function of $J$ and $U$ on a square lattice with $N=121$. The optimal configurations obtained at points A, B and C are shown in \fig{fig_mix}(a), (b) and (c) respectively.}
\label{fig_on}
\end{figure}

\begin{figure}
\leftline{\hspace{1.2cm}(a)}
\vspace{0.1cm}
\centerline{\epsfig{figure=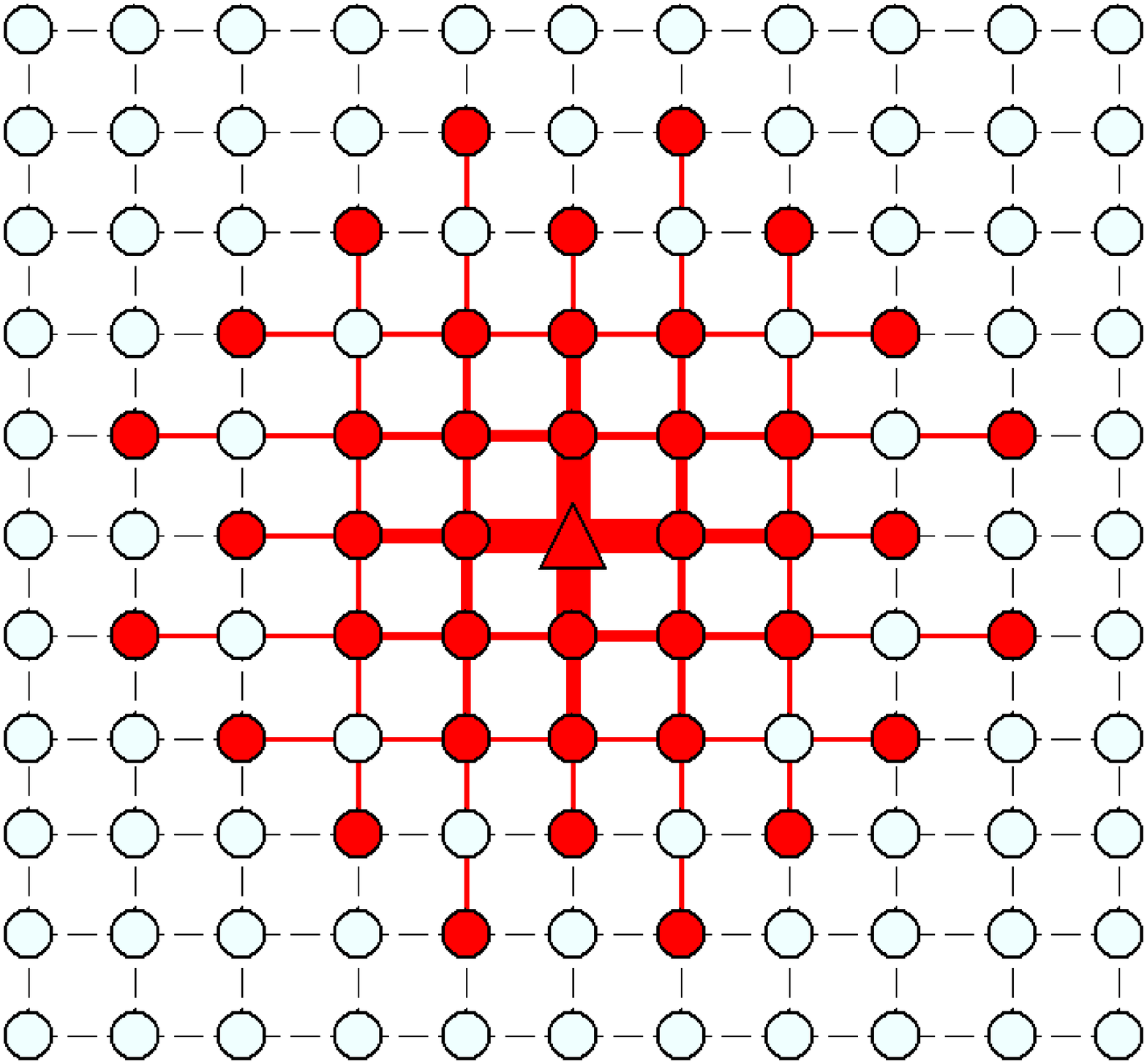, width=0.6\linewidth}}
\vspace{0.2cm}
\leftline{\hspace{1cm}(b)}
\vspace{0.1cm}
\centerline{\epsfig{figure=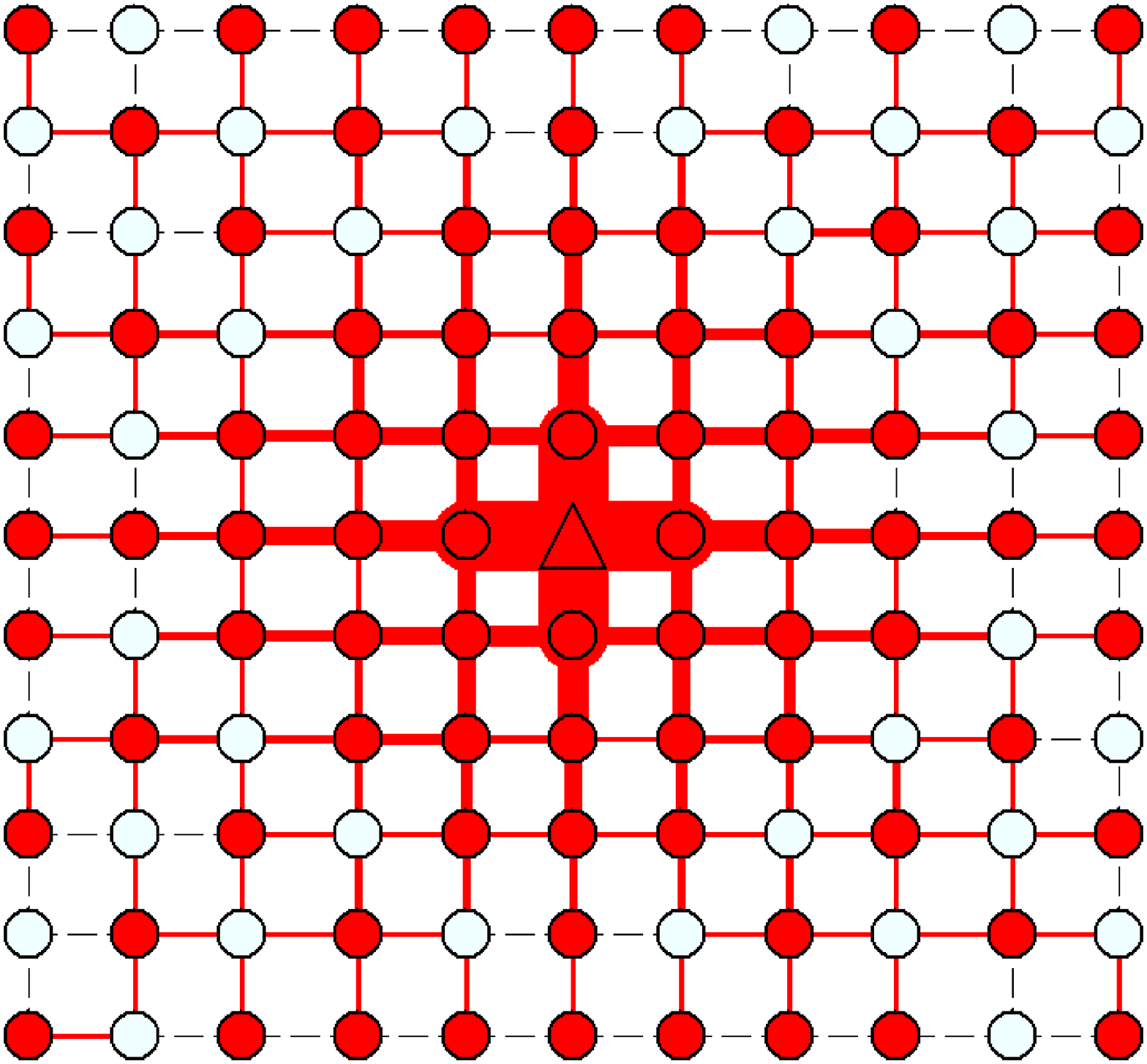, width=0.6\linewidth}}
\vspace{0.2cm}
\leftline{\hspace{1cm}(c)}
\vspace{0.1cm}
\centerline{\epsfig{figure=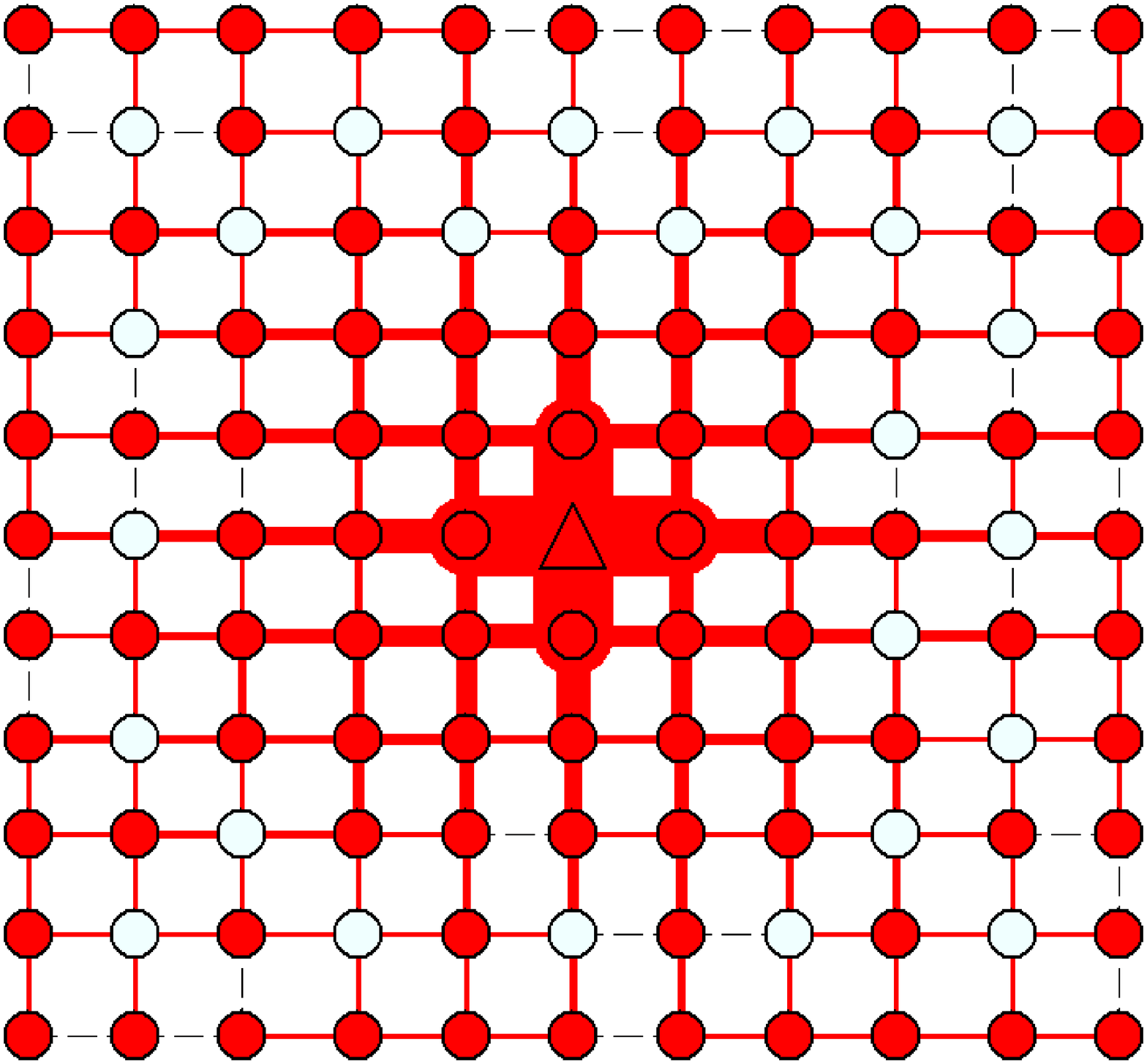, width=0.6\linewidth}}
\caption{
 (Color online) The optimized  node configuration in the mixed phase on a square lattice of N = 121 with (a) $J = 0.245$ and $U=32.6$, (b) $J = 2$, $U = 90$, (c) $J=6$ and $U = 130$.}
\label{fig_mix}
\end{figure}

We show in \fig{fig_on} the contour plot of the optimized fraction of active nodes as a function of the parameter $J$ and $U$ on square lattices with $N=121$ nodes. Although these results are obtained on a particular system size only, we will show in the next subsection that the picture is generic to different system sizes by a proper rescaling. 

We first discuss the case of $J = 0$, such that the antiferromagnetic interaction in \req{eq_H} is absent and the alternating active-idle state is not encouraged. In this case, the fraction of active nodes $\fa$ increases with idle-penalty $U$ and becomes saturated at $\fa=1$ when $U$ is larger than a certain value. It implies that more nodes become active when the idle-penalty $U$ increases. This is similar to spin models where the magnetization increases with the magnitude of external field. All the active nodes are found close to the terminal to minimize the  supply cost. In other words, there is an active-core centered at the terminal which expands with $U$ until the whole network is active. In the context of supply networks, it implies that the optimal locations of outlets are close to the warehouse when demand is low.

Next we discuss the case of non-zero antiferromagnetic coupling $J$ with $U=0$. As we can see from \fig{fig_on}, the fraction of  active nodes increases with $J$ and becomes saturated at $\fa= 0.5$. It implies that more nodes become active as $J$ increases, but unlike the previous case where there exists an active-core, a region of alternating active-idle nodes emerge around the terminal as shown in \fig{fig_ex}(b). When $J$ is larger than a certain threshold, the alternating active-idle configuration spans the whole lattice and half of nodes become active; $\fa$ saturates at 0.5. This corresponds to the case where outlets cover the whole network but are installed in every other location.

In the cases with non-zero $U$ and $J$, we note an \emph{all-active phase} with $\fa$ = 1 when $U$ is large, corresponding to a comprehensive coverage as well as a magnetized phase in a spin system where all spins align with the magnetic field. On the other hand, when $J$ is large we expect an \emph{active-idle phase} emerges where the alternating active-idle pattern spans the whole network, similar to an antiferromagnet. The region with $\fa=0.5$ does not necessarily identify the active-idle phase, since configurations other than the antiferromagnetic pattern may also have $N/2$ active nodes, for instance, a core with $N/2$ active nodes centered at the terminal. We thus measure  $\fai$, the fraction of edges which connect an active and an idle node,  and identify the antiferromagnetic phase to be the region with $\fai=1$. As we can see in \fig{fig_on}, the active-idle phase emerges as a triangular regime at large $J$.

To reveal the configurations outside the magnetized and antiferromagnetic regimes, we measure $\fan$ and $\fon$, the fraction of active nodes with at least one active neighbor and the fraction of nodes where all neighbors are in the opposite state. The four nodes at the corners are not counted in $\fon$ since they may satisfy its definition even when the antiferromagnetic interaction is absent; for instance, the four idle nodes at the corners of the configuration in \fig{fig_simple}(f). We can then identify the \emph{mixed phase} as the regime with non-zero $\fan$ and $\fon$. In the mixed phase, we generally observe an active core centered at the terminal, surrounded by an area of alternating active-idle nodes, and possibly followed by an idle area depending on the value of $U$ and $J$. For instance, such a configuration is observed in the optimized states in \fig{fig_mix}(a) and (b) which are obtained by $U$ and $J$ at point A and B in the mixed phase (see \fig{fig_on}). Other configurations emerge at larger $U$ and $J$, such as the state in \fig{fig_mix}(c) which shows an active-idle ring surrounded by a fully active boundary. The reason for such a configuration is the weaker antiferromagnetic interaction due to fewer neighbors, i.e. less redundancy, on the boundary compared to the interior nodes, which results in a preference of an active boundary over an active-idle chain along it. We remark that this configuration is suitable for surveillance network where cameras are mainly installed more on the boundary than in the interior.

Finally, the unshaded regime outside the magnetized, antiferromagnetic and the mixed phase in \fig{fig_on} is characterized by $\fa<1$ and $\fai<1$, as well as either $\fan=0$ or $\fon=0$, indicating the presence of either a magnetized or antiferromagnetic core which does not span the whole network. We thus call this regime the \emph{simple core phase}. As we have discussed, the cases of $J=0$ and $U=0$ are characterized by an active core and an active-idle core respectively, which suggests that the vertical unshaded regime near $J=0$ and to the left of the mixed phase is contributed by configurations with an active core, while the unshaded regime to the right of the mixed phase is contributed by an active-idle core.

\begin{figure}
\centerline{\epsfig{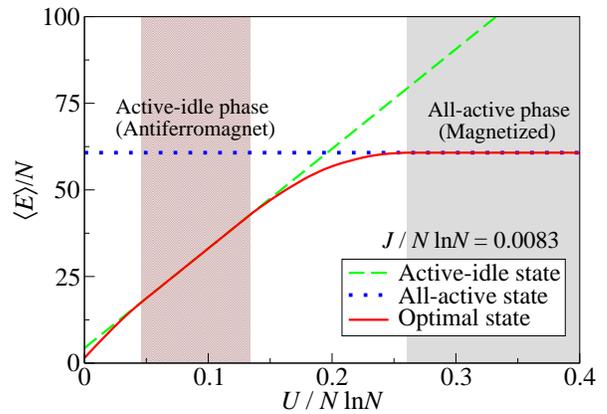}}
\caption{
(Color online) The dependence of the energy on $U$ of the all-active, the active-idle and the optimal states of a square lattice  with $N=121$ and $J/N \ln N=0.0083$.}
\label{fig_energy}
\end{figure}

With the extensive regions of the all-active and active-idle states in \fig{fig_on}, it may be misleading to conclude that optimal solutions in the mixed phase which have a more complex configurations do not result in significant gains over all-active or active-idle state. Figure~\ref{fig_energy} shows that the energy of the optimal state is significantly lower than the all-active and active-idle states where $\fa$ lies between 0.5 and 1. From the application point of view, the mixed phase is the most interesting regime, since the optimal state consists of both the all-active and the active-idle regions, but the determination of their boundary is a non-trivial task.

\subsubsection{The scaling with system size}
\label{sec_size_sq}

We further investigate the scaling behaviors of the system as a function of  size $N$. We will first identify for the cases of $U=0$ and $J=0$ a scaling factor for $U$ and $J$ which results in a data collapse of $\fa$ for different $N$. We will then examine the scaling in the mixed phase when both $U$ and $J$ are non-zero.

\begin{figure}
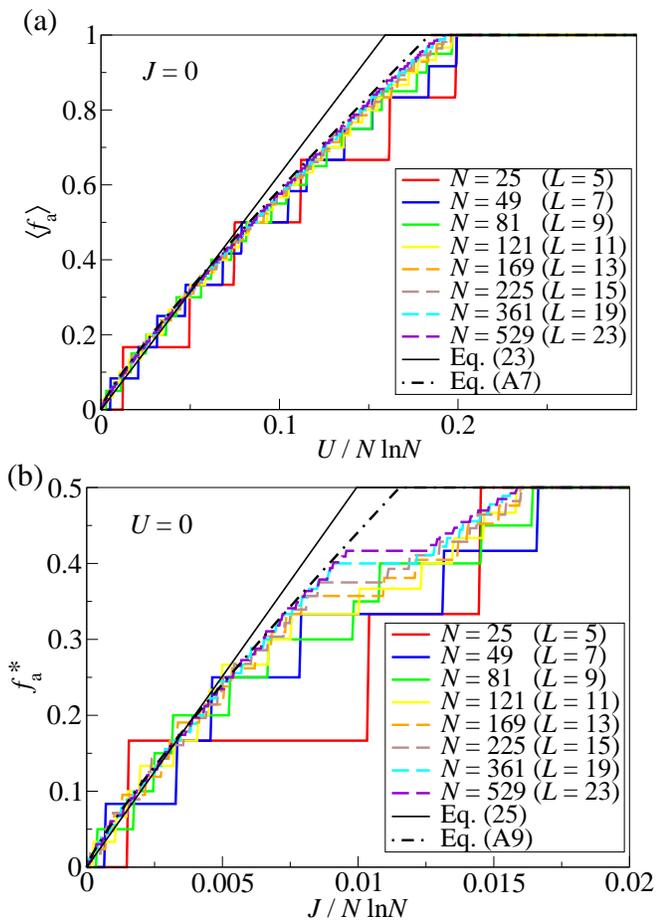

\centerline{\epsfig{figure=square_J0_4.eps, width=0.95\linewidth}}
\centerline{\epsfig{figure=square_U0_3.eps, width=\linewidth}}
\caption{
 (Color online) (a) The fraction of active nodes $\avg{\fa}$ as a function of $U/N \ln N$ with $J = 0$ and various $N$ from $N = 25$ to $N = 529$. Results are averaged over 10 realizations with different set of random biases. Analytical approximation Eqs. (\ref{eq_cont}) and (A7) with $N=529$ are shown. (b) The fraction of active nodes $\fa^*$ as a function of $J/N \ln N$ with $U=0$. Results are obtained in the minimum energy instance in 10 different instances. Analytical approximation Eqs. (\ref{eq_contJ}) and (A9) with $N=529$ are shown.}
\label{fig_fonN}
\end{figure}

To get the correct scaling factor of $U$ when $J=0$, we  adopt  the continuum approximation to compute the  supply cost. We denote the distance of the furthest active node from the terminal as $d_{\max}$. Noting that the dominant contribution to the supply cost comes from the region near the center, we can assume that the flows are isotropic and equal to $\fa N/(2\pi r)$, so that the supply cost $\sum_{ij}I_{ij}^2$ can be computed to the leading order as
\begin{align}
\sum_{(ij)}I_{ij}^2\approx\int_\epsilon^{d_{\max}}\left(\frac{\fa N}{2\pi r}\right)^2 2\pi rdr
=\frac{\fa^2 N^2}{2\pi}\ln \frac{d_{\max}}{\epsilon}.
\end{align}
The cutoff distance $\epsilon$ is of the order 1 and can be approximated as 1 to the leading order. Using $\pi d_{\max}^2=\fa N$, we obtain
\begin{align}
\label{eq_cont0}
E = (1-\fa)NU + \frac{\fa^2N^2}{4\pi}\ln\left(\fa N\right) - \frac{\fa^2N^2}{4\pi}\ln\pi,
\end{align}
which gives rise, to the leading order,
\begin{align}
\label{eq_cont}
\fa = 2\pi\frac{U}{N\ln N}.
\end{align}
This shows that the energy scale for achieving extensive coverage is $U \sim N\ln N$ when $J=0$.  As shown in \fig{fig_fonN}(a), for individual system size $N$, the fraction of active nodes $\fa$ increases with $U$, characterized by functions in the form of staircase due to the discreteness of the lattice. The larger the system, the smaller the step size and the more the steps until $\fa$ saturates at 1. Due to the different step sizes, $\fa$ as a function of $U/(N \ln N)$ for various $N$ do not overlap perfectly on one another, but interestingly there exhibits a data collapse with individual staircases collapsing on a common trend. We remark that the un-scaled counterpart of \fig{fig_fonN}(a), i.e. the number of active nodes $N_a$ as a function of un-scaled $U$ also collapses well for various $N$ as shown in \fig{fig_step}. Nevertheless, the optimized $\fa$ in \fig{fig_fonN}(a) does not match well with \req{eq_cont} when $\fa$ approaches $1$. This is due to corrections to the supply cost of the order $N^2$, which is significant compared with the leading order of $N^2 \ln N$ for the system sizes we simulated. These corrections are worked out in Appendix A and  the result of Eq. (A7) is plotted in \fig{fig_fonN}(a), producing a much improved agreement with the simulation results.

Next, we examine the case of increasing $J$ with $U=0$. In this case, we expect a core of alternating active-idle nodes emerges and  expands with $J$, similar to \fig{fig_simple} except that the core  is replaced by an alternating active-idle configuration. As shown in \fig{fig_fonN}(b), $\fa$ with different $N$ collapse well when plotted as a function of the rescaled $J/(N  \ln N)$, similar to the case of increasing $U$ with $J = 0$ in \fig{fig_fonN}(a). The reason for showing $\fa^*$, the fraction of active nodes obtained in the instance with lowest energy in 10 different realizations, will be explained in the next subsection. These results suggest  a continuum approximation similar to Eqs. (\ref{eq_cont0}) and (\ref{eq_cont}) is valid even with the alternating active-idle node configuration. Hence we model the configuration by a circular core of active-idle configuration with radius $R$, and all nodes beyond are idle. The total energy becomes
\begin{align}
E = \frac{\fa^2N^2}{4\pi}\left(\ln\left(\fa N\right)-\ln\frac{\pi}{2}+\pi-\frac{3}{2}\right) + 2JN(1-4\fa),
\end{align}
yielding, to the leading order,
\begin{align}
\label{eq_contJ}
\fa = 16\pi\frac{J}{N\ln N}.
\end{align}
 Alternatively, one can substitute $U =8J$ in \req{eq_cont} to obtain \req{eq_contJ}, since the energy gain by activating one node is generally $8J$ in the case with $U=0$. As shown in \fig{fig_fonN}(b), there is an excellent agreement between the approximation and the simulation results. On the other hand, plateaus of $\fa$ are observed at intermediate values of $J$. It is indeed an interesting phenomenon related to the lattice boundary --- since boundary nodes have fewer neighbors, their benefit from satisfying the antiferromagnetic interaction is smaller, and a larger value of $J$ is required to activate them. This enhances the stability for configurations where the alternating core spans the whole network except the boundary (see \fig{fig_ex}(b) for instance), which leads to a plateau in $\fa$ until $J$ is sufficiently large to activate boundary nodes.

\begin{figure}
\centerline{\epsfig{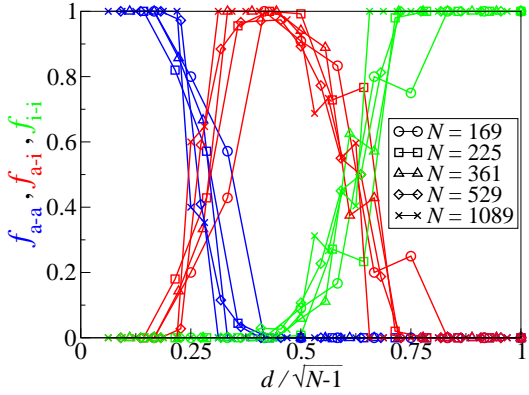}}
\caption{
 (Color online) The fraction $\faa$, $\fai$ and $\fii$ of edges connecting two active nodes, one active and one idle node, and two idle nodes as a function of rescaled distance $d/(\sqrt{N}-1)$.}
\label{fig_rescaleBond}
\end{figure}

To examine the scaling of $N\ln N$ when both $U$ and $J$ are non-zero, we compare the microscopic features of systems with different size $N$ but the same value of $U/(N\ln N)$ and $J/(N\ln N)$. For a quantitative comparison, we measure $\faa(d)$, $\fai(d)$ and $\fii(d)$, respectively the fraction of edges which connect two active nodes, one active and idle node, and two idle nodes at a Manhattan distance $d$ from the terminal. We obtain these results for systems of different $N$ but at the same values of $U/(N\ln N)$ and $J/(N\ln N)$ indicated by point A in the mixed phase of \fig{fig_on}; its corresponding optimized configuration on the $N=121$ lattice is shown in \fig{fig_mix}(a).

As we can see in \fig{fig_rescaleBond}, the results of $\faa(d)$, $\fai(d)$ and $\fii(d)$ as a function of $d/(\sqrt{N}-1)$ for different system size collapse well. The results indicate the presence of an all-active core in the neighborhood of the terminal, surrounded by an active-idle region, which is in turn surrounded by a region of idle nodes near the boundary in the various systems examined. It suggests that systems of different sizes behave similarly at the same values of $U/(N\ln N)$ and $J/(N\ln N)$, and the scaling factor $1/(N\ln N)$ is appropriate when both $U$ and $J$ are non-zero. The data collapse shown in \fig{fig_fonN} and \fig{fig_rescaleBond} suggest that phase diagrams similar to \fig{fig_on} can be identified for other system sizes related by the scaling factor $1/(N\ln N)$.

\subsubsection{Sub-optimal metastable states}

\begin{figure}
\centerline{\epsfig{figure=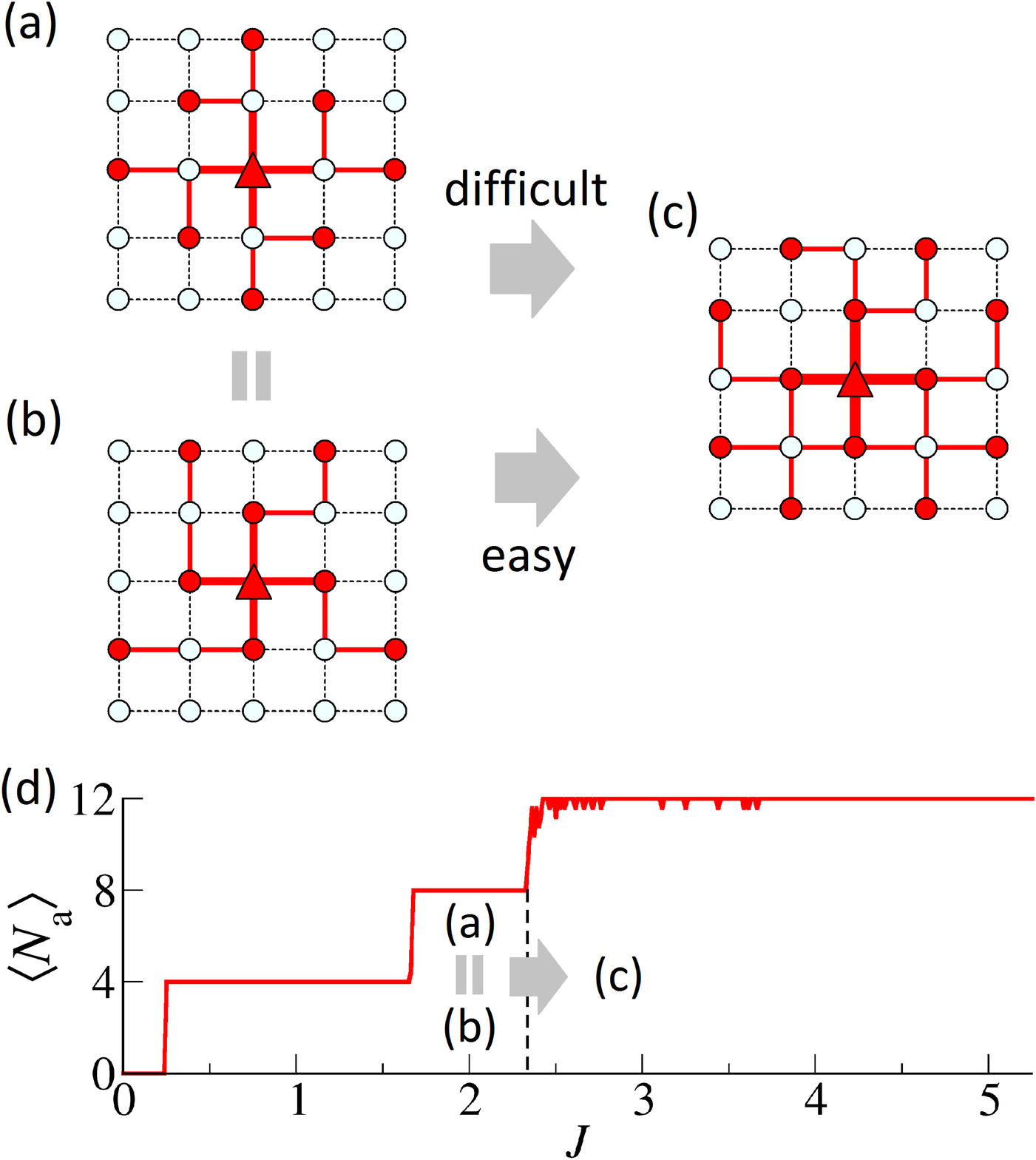, width=0.9\linewidth}}
\caption{
(Color online) (a,b) Two degenerate optimal states of a lattice with $N = 25$ in the range $0.083 < J < 1.11$ with $U=0$; (c) the corresponding optimal state for $J>1.11$; (d) the number of active nodes $\avg{N_a}$ as a function of $J$ averaged over 10 realizations.}
\label{fig_simpleJ}
\end{figure}

We found that sub-optimal solutions are sometimes obtained for some values of $J$ when $U=0$. This is in contrast to the simpler case of $J=0$, where the optimal solution is always found by the algorithm (for instance, see \fig{fig_step}). These result in a rough curve of the average number of active nodes $N_a$ with $J>1.11$, as shown in \fig{fig_simpleJ}(d). The reason can be illustrated by the example shown in \fig{fig_simpleJ}(a)-(c). In the range $0.83<J<1.11$, both configuration (a) and (b) in \fig{fig_simpleJ} have the lowest energy; when $J>1.11$, configuration (c) becomes the ground state. The transition from configuration (b) to (c) is easy, since one can merely activate the four nodes next to the four corners in (b); the transition from (a) to (c) requires an extensive change of the variables for the whole network in (a), and is difficult. Configuration (a) hence becomes a local minimum separated by high energy barriers when $J>1.11$, and the algorithm arrives at this sub-optimal solution if the initial messages fall in its basin of attraction. The competition between the two antiferromagnetic configurations with opposite staggered magnetizations are crucial for the emergence of these metastable states.

One can easily eliminate this algorithmic deficiency by studying only the minimum energy state in multiple realizations. As shown in \fig{fig_fonN}(b), the roughness observed in \fig{fig_simpleJ}(d) is eliminated by measuring $\fa^*$, the fraction of active nodes in the instance with the lowest energy in 10 realizations. This is the reason for showing $\fa^*$ instead of $\avg{\fa}$ in \fig{fig_fonN}(b) in the case with $U=0$. 

\subsection{Random regular graph}
\label{sec_rrg}

We continue to examine the optimized configuration of active nodes by running our algorithm on random regular graphs, i.e. randomly connected networks with uniform degree $K$. In general, only loops with size $O(\ln N)$ exist in random regular graphs, which makes the assumption of independent neighbors in \req{eq_cavity} more justified than in the square lattice. We thus expect a better algorithmic convergence, but results show that the algorithm does not converge in a broad parameter regime. For instance, as shown in \fig{fig_rsb}, the ratio of convergent instances $\fcon$ significantly decreases when $J>0.095$ with $U=0$. This is due to the presence of loops with an odd number of links, which are frustrated for antiferromagnetic interactions, in contrast with square lattices whose loops consist of even number of links, accounting for the excellent algorithmic convergence.

\subsubsection{The various phases}

\begin{figure}
\centerline{\epsfig{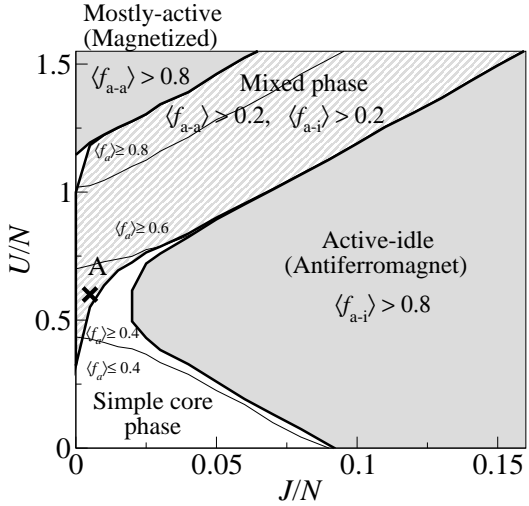}}
\caption{
Phase diagram and the fraction of active nodes as a function of $J$ and $U$ on random regular graphs with $N=50$ and node degree $K=3$. Results are averaged over 1000 realizations.}
\label{fig_rrg_phase}
\end{figure}

Unlike square lattices where a unique optimized energy is found at each specific value of $U$ and $J$, there is a non-zero variance of optimized energy obtained from an ensemble of finite-size random regular graphs at each $U$ and $J$. We therefore observed smoothened contour lines of $\avg{\fa}$ in \fig{fig_rrg_phase} compared to the zigzag contour lines in \fig{fig_on} of square lattices. In addition,
due to the presence of loops with an odd number of links, $\avg{\fai}$ over realizations is always less than 1 even in the active-idle phase. We have thus chosen a reasonable threshold on $\avg{\fai}$ and $\avg{\faa}$ as an identification criteria of the active-idle and the mixed phase. To identify the magnetized phase, we note that although $\faa$ approaches 1 for sufficiently large $U$, it is nevertheless close to 1 for an extensive range of $U$. We have thus applied similar threshold on $\avg{\faa}$ to identify the magnetized phase on typical instances.

Despite these slight differences in the identification of phases, random regular graphs and square lattices have similar behavior as a function of $U$ and $J$, as can be seen from the similarity of their corresponding phase diagram. As we can see in \fig{fig_rrg_phase}, random regular graphs exhibit similar phase diagram as that of square lattices, and are dominated by mostly-active magnetized configurations at large value of $U$ characterized by $\avg{\faa}>0.8$, active-idle antiferromagnetic configurations at large value of $J$ characterized by $\avg{\fai}>0.8$, as well as a mixed phase for a broad parameter space outside the magnetized and antiferromagnetic regime characterized by $\avg{\faa}>0.2$ and $\avg{\fai}>0.2$. The simple core phase is found in the unshaded region outside all these regimes.

\subsubsection{The hard computation regime}

\begin{figure}
\epsfig{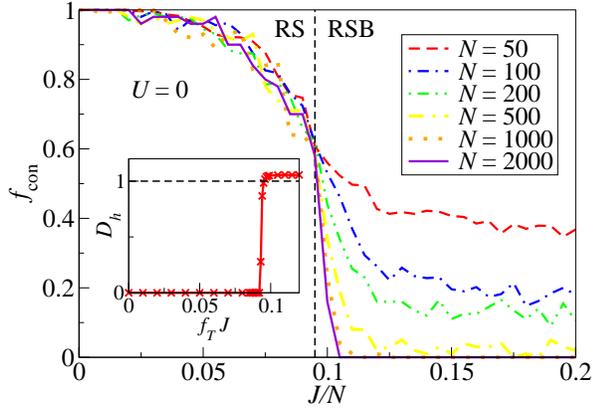}
\caption{(Color online)
The fraction of convergent instances $\fcon$ as a function $J/N$ on random regular graph with $N=50$, $K=3$ and $U=0$. Inset:  $\avg{D_h}$, the fractional change of Hamming distance given by \req{eq_dh} as a function $f_T J$, with $f_T=0.02$.
}
\label{fig_rsb}
\end{figure}

To examine the origin of the non-convergence, we note that the drop in $\fcon$ in \fig{fig_rsb} is more abrupt when $N$ increases. This suggests a phase transition at $J \approx 0.095$ and the emergence of replica symmetry breaking (RSB) in the range $J > 0.095$. RSB is a phenomenon first observed in spin glass~\cite{mezard87} which is characterized by a rugged energy landscape with numerous local minima. When one applies a message-passing algorithm to an RSB system, individual variables may fall into states from different local minima, and leads to conflicting messages and non-convergence. Algorithms which incorporate the picture of a rugged energy landscape can be derived by drawing further correspondence with spin glasses~\cite{mezard02}.

To analytically derive the RSB phase boundary, we study two identical replicated systems $\beta$ and $\gamma$ with different boundary conditions. If the  states of $\beta$ and $\gamma$ far away from the boundary are different, long-range correlations exist, suggesting the emergence of RSB. To achieve the goal, we obtain the joint probability distribution $P[E^{V}_\beta(s, I), E^{V}_\gamma(s, I)]$ by the following recursive equation:
\begin{widetext}
\begin{align}
\label{eq_cavityrsb}
&P[E^{V}_\beta(s, I), E^{V}_\gamma(s, I)] 
= f_T\delta\left\{E^{V}_\beta(s, I)-|\flow|^\alpha - J~C(1+s)\right\}
\delta\left\{E^{V}_\gamma(s, I)-|\flow|^\alpha - J~C(1+s)\right\}
\nonumber\\
&\qquad+(1-f_T)
\sum_{k=1}^\infty \frac{k\rho(k)}{\avg{k}}  \prod_{j=1}^{k-1} \left[\int d E^{V}_{j,\beta}(s_j, I_j) d E^{V}_{j,\gamma}(s_j, I_j) ~P[E^{V}_{j,\beta}(s_j, I_j), E^{V}_{j,\gamma}(s_j, I_j)] \right]
\nonumber\\
&\qquad\times\delta\left\{
E^V_{\beta}(s, I) - \cM\left[s, I; \{E^V_{j,\beta}(s_j, I_j)\}_{1\le j\le k-1}\right] + \min_{s', I'}\cM\left[s', I'; \{E^V_{j,\beta}(s_j, I_j)\}_{1\le j\le k-1}\right]\right\}
\nonumber\\
&\qquad\times\delta\left\{
E^V_{\gamma}(s, I) - \cM\left[s, I; \{E^V_{j,\gamma}(s_j, I_j)\}_{1\le j\le k-1}\right] + \min_{s', I'}\cM\left[s', I'; \{E^V_{j,\gamma}(s_j, I_j)\}_{1\le j\le k-1}\right]\right\}
\end{align}
\end{widetext}
where $f_T$ is the fraction of terminals in the infinite system assumed by the equation. Since we usually assign only one terminal in simulations, $f_T$ should be set to the $1/N_{\rm sim}$ for comparison, where $N_{\rm sim}$ is the simulated system size. The last two lines   represent two identical operations, one for replica $\beta$ and the other for replica $\gamma$. Hence the expression characterizes two replicated systems with identical quenched disorders. A simple initial condition of uniform $P[E^{V}_\beta(s, I), E^{V}_\gamma(s, I)]$  can already distinguish the RS/RSB behaviors: if one obtains a stable solution of $P[E^{V}_\beta(s, I), E^{V}_\gamma(s, I)]=0$ everywhere except on the diagonal $E^{V}_\beta(s, I) = E^{V}_\gamma(s, I)$, then the system has a unique global minimum and is characterized by the RS ansatz; on the other hand, if one obtains a stable solution of $P[E^{V}_\beta(s, I), E^{V}_\gamma(s, I)]\neq 0$ for the off-diagonal domain, then the system is likely to be characterized by the RSB ansatz.

\begin{figure}
\centerline{\epsfig{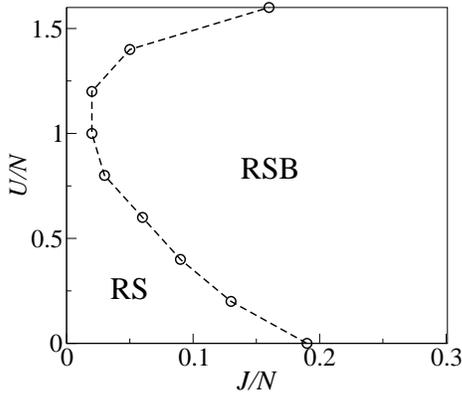}}
\caption{
A phase diagram which shows the RS/RSB boundary obtained by monitoring $D_h$. Results are obtained with $f_T=0.02$ and $K=3$. 
}
\label{fig_phase}
\end{figure}

Nevertheless, the stable solution of $P[E^{V}_\beta(s, I), E^{V}_\gamma(s, I)]$ is crucially dependent on the choice of initial condition in the RSB regime and is difficult to obtain. Instead of computing all the stable solutions of $P[E^{V}_\beta(s, I), E^{V}_\gamma(s, I)]$ by starting with all possible initial conditions, we introduce
\begin{align}
d_\beta(I) = E^{V}_{\beta}(1, I)-E^{V}_{\beta}(-1, I),
\end{align}
which characterizes the energy difference between the cases when node $j$ is active ($s_j=1$) and idle ($s_j=-1$) in system $\beta$. We compute both $d_\beta(I)$ and $d_\gamma(I)$ and monitor the change in Hamming distance between $\beta$ and $\gamma$ by 
\begin{align}
\label{eq_dh}
D_h = \frac{ \sqrt{\sum_{I}\left[d_\beta(I)-d_\gamma(I)\right]^2 } }
{ \frac{1}{k-1}\sum_{j=1}^{k-1}\sqrt{\sum_{I}\left[d_{j, \beta}(I)-d_{j, \gamma}(I)\right]^2 } }
\end{align}
during the iteration of \req{eq_cavityrsb}. The quantity $D_h$ is thus the fractional increase of Hamming distance during each iteration. We then average $D_h$ over all nodes in the infinite tree generated by the population dynamics. If $\avg{D_h}<1$, the difference between $\beta$ and $\gamma$ decreases and is eventually washed out; the system is described by the RS ansatz. If $\avg{D_h}>1$, the difference increases and the system is described by the RSB ansatz. The quantity $D_h$ obtained by $f_T=0.02$ is shown in the inset of \fig{fig_rsb}, which indicates that $J/N\gtrsim 0.095$ is characterized by the RSB ansatz at $U=0$. The phase diagram \fig{fig_phase} is drawn by identifying the parameter regime with $\avg{D_h}>1$, which indicates the RS/RSB phase boundary. It shows that the RSB regime covers both the active-idle phase and the mixed phase.

\subsubsection{The scaling with system size}

\begin{figure}
\centerline{\epsfig{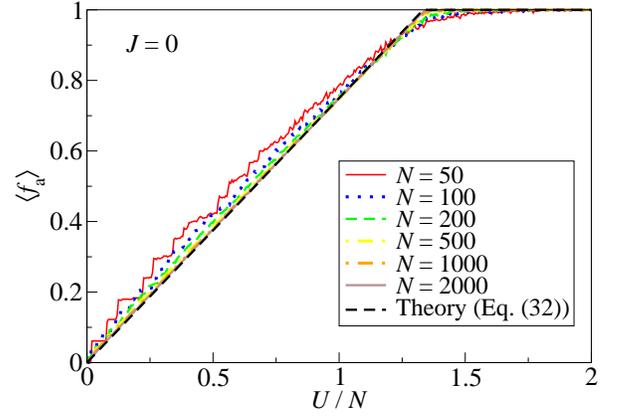}}
\caption{(Color online)
The fraction of active nodes $\fa$ on random regular graphs as a function of $U/N$ with $J=0$, $K=3$ and various $N$ from $N=50$ to $N=2000$. The theoretical predictions are obtained from \req{eq_fon_rrg}.}
\label{fig_fonN2}
\end{figure}

As in the case of square lattices, we examine the scaling of the fraction of active nodes with increasing system size $N$. To find the correct scaling, we first consider the case of $J=0$ and obtain an approximation for the  supply cost $\sum_{(ij)}I^2_{ij}$ as a function of idle penalty $U$. We  define the distance $d$ between edges/nodes and the terminal  to be the minimum number of hops along a connected path between them. In this case, we assume that the random regular graphs can be well described by a tree structure such that the number of edges and nodes at a distance $d$ from the terminal is $K(K-1)^d$, the total number of nodes within a distance $d$ is $K[1+(K-1)+\cdots+(K-1)^{d-1}]=K[(K-1)^d-1]/[K-2]$. Again, we define $d_{\max}$ to be the distance of the farthest active node from the terminal, such that
\begin{align}
\label{eq_fonN_rrg}
\fa N\approx \frac{K[(K-1)^{d_{\max}}-1]}{K-2}.
\end{align}
The  supply cost can be approximated by
\begin{align}
\label{eq_approx_current2}
\sum_{(ij)}I_{ij}^2 \approx &
\sum_{d=1}^{d_{\max}}\frac{\left(\fa N\!-\frac{K[(K\!-\!1)^{d}-1]}{K\!-\!2}\right)^2}{K(K\!-\!1)^{d}}
\nonumber\\
\approx&\frac{\left(\fa N\right)^2}{K}\sum_{d=1}^{d_{\max}}\frac{1}{(K-1)^d},
\nonumber\\
\approx&\frac{(\fa N)^2(K-1)}{K(K-2)}.
\end{align}
To arrive at  the last two lines, we have used the fact that $K[(K-1)^d-1]/[K-2]\ll \fa N$ when $d$ is small, and the contribution from terms with large $d$ is negligible. Using the above results, we can approximate the Hamiltonian in  \req{eq_H} by
\begin{align}
E = (1-\fa)NU + \frac{(\fa N)^2(K-1)}{K(K-2)}.
\end{align}
Minimizing E with respect to $\fa$, we obtain
\begin{align}
\label{eq_fon_rrg}
\fa = \frac{U}{N}\frac{K(K-2)}{2(K-1)}.
\end{align}

As shown in \fig{fig_fonN2}, the simulated results of $\fa$ for different system sizes collapse well when plotted as a function of $U/N$. Compared with square lattices in which $\fa$ scales as $U/(N\ln N)$, this reveals a crucial difference between two-dimensional lattices and random graphs. This result also suggests that in the large $N$ limit, covering a random graph is more efficient than covering a square lattice since the costs $U$ and $J$ scale linearly with $N$. From \req{eq_H}, we may interpret $U$ to be the cost increase when one additional node is activated. This implies that it is less costly to set up retail networks with the same fraction of outlets in random graphs. 

On the other hand, discrete jumps in $\fa$ are only observed at small values of $U$ in  \fig{fig_fonN2}, unlike \fig{fig_fonN} of square lattices where steps are observed for the whole range of $U$ and $\fa< 1$. The smoothened $\fa$ at larger values of $U$ is due to the average over realizations with different network topology. Nevertheless, the simulated $\fa$ approaches the analytical result \req{eq_fon_rrg} as $N$ increases. Data collapse of $\fa$ is also observed in the case of increasing $J$ with $U=0$.

\begin{figure}
\centerline{\epsfig{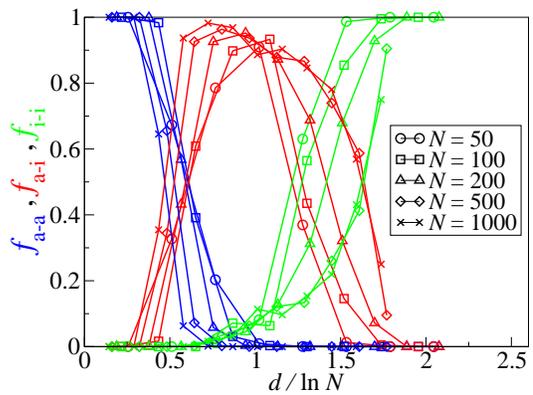}}
\caption{
 (Color online) The fraction $\faa$, $\fai$ and $\fii$ of edges connecting two active nodes, one active and one idle node, and two idle nodes as a function of rescaled distance $d/\ln N$. Results are averaged over more than 300 converged instances for $N=50, 100, 200$ and a smaller number of instance for $N=500, 1000$ due to lower convergence rate.}
\label{fig_rrg_rescaleBond}
\end{figure}

As in the case of square lattices, we examine the scaling in the mixed phase by comparing $\faa(d)$, $\fai(d)$ and $\fii(d)$ obtained from systems of different $N$ but at the same value of $U/N$ and $J/N$. As shown in \fig{fig_rrg_rescaleBond}, the results obtained by the values of $U/N$ and $J/N$ at point A of the phase diagram \fig{fig_rrg_phase} indicates the presence of an active core near the terminal, followed by an active-idle region, and then by a region with higher density of idle nodes. Similar configurations are found in square lattices (see \fig{fig_rescaleBond}) at a similar point A in the phase diagram \fig{fig_on}. Although the data do not collapse well in \fig{fig_rrg_rescaleBond} for small system sizes, we observe a better collapse for larger system sizes, which suggests that the scaling factor $1/N$ is appropriate for cases with both $U$ and $J$ non-zero and at large values of $N$.

\section{Conclusion}

We have studied the problem of optimal facility location by mapping it to a model of interacting spins, and applied statistical physics from the studies of disordered systems to elucidate the effects of competing considerations on the system behavior. We found that in a large parameter region, a magnetized domain exists in the core region, surrounded by an antiferromagnetic domain, which is in turn surrounded by an idle domain. Besides the coexistence of these three patterns, phases dominated by each pattern can be obtained by tuning the idle penalty and the antiferromagnetic coupling. The scaling of parameters with system size agrees with the continuum approximation in two dimensions and the tree approximation in random graphs.

The above results agree with the intuition that transportation costs favor the selection of active nodes in the vicinity of the center. However, they also show that by introducing antiferromagnetic couplings with appropriate strengths, the active nodes can spread over the arena, enhancing the coverage. Combined with the message-passing algorithm, this provides a simple decentralized control measure in such applications as sensor networks in which it is desirable to collect information spread over the entire arena.

To optimize the cost function, we derived a readily applicable algorithm which balances the need for coverage and energy-saving. Compared with traditional mixed integer programming, this algorithm scales favorably with system size. In addition to the conventional task of locating facilities, our algorithm identifies the optimal quantity of demand and optimizes individual paths. The local expansion approach suggested in the message updates also shed light on similar simplifications in other problems. Compared with traditional approaches of global optimization, the message-passing algorithm is useful in applications to large networks and dynamically evolving networks. In large networks, decentralized approaches have advantages over global optimization, which scales up rapidly with system size and requires heavy overhead. In dynamically evolving networks such as mobile sensor networks, the changing topology implies that global optimization approaches are not temporally relevant. Rather, locally optimal solutions should be able to cope with the evolving situations.

On the other hand, due to frustrations caused by the competition between coverage and supply cost, a transition between easy and hard computation regimes is observed. The hard computation regime coincides with the antiferromagnetic phase and the mixed phase, where metastable states are prevalent. Convergence of the message-passing approach is affected. Further computational techniques such as decimation~\cite{mezard02} or reinforcement~\cite{chavas05} may be introduced to alleviate the problem

\section{Acknowledgement}

We thank David Saad, Jack Raymond, Alan Fung, Hugh Wang and Paul Choi for fruitful discussions. This work is supported by Research Grants Council of Hong Kong (605010, 604512 and 605813), EU FET project STAMINA (FP7-265496) and Royal Society Exchange Grant IE110151. 
 
\appendix

\section{The Total Cost in the Continuum Approximation for $J=0$} 

We first calculate the supply cost to the order $N^2$. Assuming that the flows are isotropic, the flux (i.e., the flow per unit cross sectional length) at distance $r$ from the center is given by
\begin{align}
J(r) = \frac{\pi(d^2_{\max}-r^2)}{2\pi r}.
\end{align}
Hence the supply cost is calculated to be
\begin{align}
\label{eq_cont_supplyE}
E_{supply} = \int^{d_{\max}}_\epsilon J^2(r) 2\pi r dr = \frac{\pi d_{\max}^4}{2}\left(\ln\frac{d_{\max}}{\epsilon}-\frac{3}{4}\right),
\end{align}
where we have assumed $\epsilon\ll d_{\max}$. The result is rather sensitive to the cutoff distance $\epsilon$ in two-dimension lattices. We derive its value by noting that the flux can be considered as the gradient of the chemical potential. By requiring the potential difference between the center and its neighbor to be equal to the flow in their link, we can obtain a self-consistent condition for the cutoff. Specifically, we let $J(r) = -d\mu(r)/dr$, so that
\begin{align}
\mu(r) = -\int J(r)dr = -\frac{d^2_{\max}}{2}\ln r + \frac{r^2}{4}.
\end{align}
The flow in a link crossing the boundary is given by
\begin{align}
\mu(0)-\mu(1) = -\frac{d^2_{\max}}{2}\ln\epsilon - \frac{1}{4} = \frac{\fa N}{4}.
\end{align}
Using $\pi d^2_{\max} = \fa N$, we arrive at $\epsilon = \exp(-\pi/2)$, and \req{eq_cont_supplyE} becomes
\begin{align}
E_{\rm supply} = \frac{\fa^2N^2}{4\pi}\left(\ln\left(\fa N\right)-\ln\pi+\pi-\frac{3}{2}\right).
\end{align}
The total cost becomes
\begin{align}
E =& (1-\fa)NU + \frac{\fa^2N^2}{4\pi}\left(\ln\left(\fa N\right)-\ln\pi+\pi-\frac{3}{2}\right) 
\nonumber\\
&+  \frac{\fa^2N^2}{4\pi}\ln \fa.
\end{align}
The optimal fraction of active nodes satisfies the condition
\begin{align}
\fa\left(\ln N + \ln \fa - \ln\pi+\pi-1\right) = \frac{2\pi U}{N}
\end{align}

Similarly, one can derive the fraction of active nodes as a function of $J$ when $U=0$. Since only half of the nodes are active within the distance $d_{\max}$ in this case, $\pi d_{\max}^2 = 2f_aN$ and the flux is given by
\begin{align}
J(r) = \frac{1}{2}\frac{\pi(d_{\max}^2-r^2)}{2\pi r}.
\end{align}
Following the line of the above calculation, we arrive at
\begin{align}
\fa\left(\ln N + \ln \fa - \ln\frac{\pi}{2}+\pi-1\right) = \frac{16\pi J}{N}.
\end{align}



\begin{thebibliography}{10}

\bibitem{rangwala06}
S.~Rangwala, R.~Gummadi, R.~Govindan and K.~Psounis,
\newblock Computer Communication Review {\bf 36}, 63 (2006).

\bibitem{megiddo81}
N.~Megiddo, E.~Zemel and S.~L. Hakimi,
\newblock SIAM Journal on Algebraic and Discrete Methods {\bf 4}, 253 (1981).

\bibitem{vasan10}
A.~Vasan and S.~P. Simonovic,
\newblock Journal of Water Resources Planning and Management {\bf 136}, 279
  (2010).

\bibitem{ghosh83}
A.~Ghosh and C.~S. Craig,
\newblock Journal of Marketing {\bf 47}, 56 (1983).

\bibitem{berman90}
O.~Berman, R.~C. Larson and N.~Fouska,
\newblock MIT Operations Research Center Working Paper OR 231-90  (1990).

\bibitem{revelle77}
C.~ReVelle {\em et~al.},
\newblock Health Services Res. (Summer 1977) {\bf 12}, 129 (1977).

\bibitem{alkaraki04}
J.~N. Al-Karaki and A.~E. Kamal,
\newblock IEEE Wireless Communications {\bf 11}, 6 (2004).

\bibitem{frey09}
H.~Frey, S.~R\:uhrup,  and I.~Stojmenovi\'c,
\newblock {\em Guide to Wireless Sensor Networks (edited by S. Misra, S. C.
  Misra, and I. Woungang)} (Springer, London, United Kingdom, 2009).

\bibitem{seward88}
D.~Seward,
\newblock {\em Napoleon and Hitler} (Viking, New York, USA, 1988).

\bibitem{clouquer02}
T.~Clouqueur, V.~Phipatanasuphorn, P.~Ramanathan and K.~K. Saluja,
\newblock Proceedings of the 1st ACM international workshop on wireless sensor
  networks and applications , 42  (2002).

\bibitem{zou03}
Y.~Zou and K.~Chakrabarty,
\newblock 22nd Annual Joint Conference of the IEEE Computer and Communications, 1293  (2003).

\bibitem{plastria01}
F.~Plastria,
\newblock European Journal of Operational Research {\bf 129}, 461 (2001).

\bibitem{wong08}
K.~Y.~M. Wong and D.~Saad,
\newblock J. Phys. A: Math. Theor. {\bf 41}, 324023 (2008).

\bibitem{vannimenus77}
J.~Vannimenus and G.~Toulouse,
\newblock J. Phys. C {\bf 10}, L537 (1977).

\bibitem{mezard87}
M.~M\'ezard, G.~Parisi and M.~A. Virasoro,
\newblock {\em Spin Glass Theory and Beyond} (World Scientific, 1987).

\bibitem{nishimori01}
H.~Nishimori,
\newblock {\em Statistical Physics of Spin Glasses and Information Processing}
  (Oxford University Press, Oxford ,UK, 2001).

\bibitem{wong06}
K.~Y.~M. Wong and D.~Saad,
\newblock Phys. Rev. E {\bf 74}, 010104(R) (2006).

\bibitem{yeung12}
C.~H. Yeung and D.~Saad,
\newblock Phys. Rev. Lett. {\bf 108}, 208701 (2012).

\bibitem{selke88}
W.~Selke,
\newblock Phys. Rep. {\bf 170}, 213 (1988).

\bibitem{domb51}
C.~Domb and R.~B. Potts,
\newblock Proc. Roy. Soc. Lond. A {\bf 210}, 125 (1951).

\bibitem{stephenson70}
J.~Stephenson and D.~D. Betts,
\newblock Phys. Rev. B {\bf 2}, 2702 (1970).

\bibitem{bak80}
P.~Bak and J.~von Boehm,
\newblock Phys. Rev. B {\bf 21}, 5297 (1980).

\bibitem{vannimenus81}
J.~Vannimenus,
\newblock Z. Phys. B {\bf 43}, 141 (1981).

\bibitem{inawashiro83}
S.~Inawashiro, C.~J. Thompson and G.~Honda,
\newblock J. Stat. Phys. {\bf 33}, 419 (1983).

\bibitem{ganikhodjaev03}
N.~N. Ganikhodjaev, C.~H. Pah and M.~R.~B. Wahiddin,
\newblock J. Phys. A {\bf 36}, 4283 (2003).

\bibitem{pirkul98}
H. Pirkul and V. Jayaraman, Computers and Oper. Res. {\bf 25}, 869 (1998).

\bibitem{facility}
In typical facility location problems, the demands of the clients at different fixed locations are constrained to be satisfied and the locations of the facilities are selected to minimize a cost function \cite{pirkul98}. Alternatively, the cost of installing the facilities is fixed and the locations of the clients or facilities are selected to minimize the population deprived of the facilities \cite{moore82}. This brings the mathematical formulation comparable to our model, except for the presence of antiferromagnetic couplings and the quadratic transportation cost.

\bibitem{moore82}
G. C. Moore and C. ReVelle, Manage. Sci. {\bf 28}, 775 (1982).

\bibitem{cplex}
IBM ILOG CPLEX optimization studio, 
{\scriptsize\text{http://www-03.ibm.com/software/products/en/ibmilogcpleoptistud/}}
retrieved on 20th March, 2014.

\bibitem{mezard02}
M.~M\'ezard and R.~Zecchina,
\newblock Phys. Rev. E {\bf 66}, 056126 (2002).

\bibitem{yeung13c}
C.~H. Yeung,
\newblock Proceedings of 2013 IEEE International Conference on Communications Workshops, 1420 - 1424 (2013).

\bibitem{yeung13}
C.~H. Yeung, D.~Saad and K.~Y.~M. Wong,
\newblock Proc. Natl. Acad. Sci. USA  {\bf 110}, 13717-13722 (2013).

\bibitem{chavas05}
 J. Chavas, C. Furtlehner, M. M\'ezard, and R. Zecchina, J. Stat. Mech P11016 (2005).

\end{thebibliography}

\end{document}